\documentclass{jaa}
\usepackage[square]{natbib}
\usepackage[colorlinks=true, citecolor = blue]{hyperref}
\usepackage[english]{babel}
\usepackage{epsfig}
\usepackage{epstopdf}
\usepackage{psfrag}
\usepackage{color}
\usepackage{graphicx}
\usepackage{fancyhdr}
\usepackage{graphics} 
\usepackage{amsmath}
\usepackage{xspace}

\def\V2{V_2}
\def\V2ij{V_{2ij}}

\def\V{\mathcal{V}}

\usepackage{soul,xcolor}

\newcommand{\hi}{H{\sc i}\xspace}
\newcommand{\casa}{\texttt{CASA}\xspace}
\newcommand{\osc}{\texttt{OSKAR}\xspace}
\setstcolor{magenta}
\def\lsim{~\rlap{$<$}{\lower 1.0ex\hbox{$\sim$}}}

\def\gsim{~\rlap{$>$}{\lower 1.0ex\hbox{$\sim$}}}

\begin{document}\sloppy

\title{Synthetic Observations with the Square Kilometre Array (SKA) - development towards an end-to-end pipeline}

\author{Aishrila Mazumder\textsuperscript{1*}, Abhirup Datta\textsuperscript{1}, Mayuri Sathyanarayana Rao\textsuperscript{2}, Arnab Chakraborty\textsuperscript{3}, Saurabh Singh \textsuperscript{2}, Anshuman Tripathi\textsuperscript{1}, Madhurima Choudhury\textsuperscript{1,4}}

\affilOne{\textsuperscript{1}Department of Astronomy, Astrophysics \& Space Engineering, Indian Institute of Technology Indore, Indore 453552, India.\\}
\affilTwo{\textsuperscript{2}Raman Research Institute, C.V. Raman Avenue, Sadashivanagar, Bangalore 560080, India\\}
\affilThree{\textsuperscript{3}Department of Physics and McGill Space Institute, McGill University, Montreal, QC, Canada H3A 2T8.\\}
\affilFour{\textsuperscript{4}Astrophysics Research Center(ARCO), Department  of Natural Sciences, The Open University of Israel, Ra'anana 4353701\\}

\date {}
\twocolumn[{\maketitle

\corres{aishri0208@gmail.com}

\begin{abstract}
Detection of the redshifted 21-cm signal of neutral hydrogen from the Cosmic Dawn and the Epoch of Reionization is one of the final frontiers of modern observational cosmology. The inherently faint signal makes it susceptible to contamination by several sources like astrophysical foregrounds and instrumental systematics. Nevertheless, developments achieved in the recent times will combine to make signal detection possible with the upcoming Square Kilometer Array (SKA), both statistically and via tomography. This review describes an indigenously developed end-to-end pipeline that simulates sensitive interferometric observations. It mainly focuses on the requirements for \hi detection in interferometers. In its present form, it can mimic the effects of realistic point source foregrounds and systematics- calibration error and position error on 21-cm observations. The performance of the pipeline is demonstrated for test cases with 0.01\% calibration error and position error. Its performance is consistent across telescope, foreground, and signal models. The focus of the simulation pipeline during the initial stages was for EoR science. But since this is a general interferometric simulation pipeline, it will be helpful to the entire SKA user community, irrespective of the science goals.

\end{abstract} 

}]

\doinum{12.3456/s78910-011-012-3}
\artcitid{\#\#\#\#}
\volnum{000}
\year{0000}
\pgrange{1--\pageref{LastPage}}
\setcounter{page}{1}

\section{Introduction}
 The advent of extremely sensitive telescopes has resulted in massive progress towards understanding the Universe. Large galaxy redshift surveys, Cosmic Microwave Background (CMB) observations, galaxy clustering, gravitational lensing, etc., have contributed to a better understanding of the formation and evolution of the Universe has been obtained. However, while CMB probes very early phases of the Universe and galaxy surveys probe later stages, the intermediate phase, from the recombination era ($\sim$400,000 years from the Big Bang) till the time galaxy formation starts peaking ($\sim$1.5 Gyr since Big Bang) remains unexplored. This corresponds to a redshift range of 1100$\lesssim z \lesssim$3, where observations are scarce. It is expected that in the redshift range 30$\lesssim z \lesssim$15, gravitational fluctuations grew into potential wells, accumulating matter and through a series of complicated physical processes forming the first stars. This era is called the Cosmic Dawn (CD). The stars and galaxies formed during the CD ionized the post-recombination neutral intergalactic medium (IGM), resulting in the Epoch of Reionization (EoR), the last global phase transition in the history of the Universe, lasting between  15$\lesssim z \lesssim$6. Observation of the CD/EoR remains the frontier in observational cosmology. 
 
 Hydrogen, the most abundant element in the Universe, comprises 75\% of the total baryonic matter. The 21-cm line of neutral hydrogen (\hi) is one of the most reliable probes for observation and understanding of the CD/EoR \citep{loeb2007frontier, Bharadwaj_2001, Bharadwaj2001B,Bharadwaj_ali2005}. The 21-cm line from neutral hydrogen arises due to the hyperfine ``spin-flip" transition where a parallel spin electron and proton transits to a state with anti-parallel spins releasing a photon of rest-frame wavelength of $\sim$21cm (or frequency $\sim$1.4 GHz) \citep{Field1958PIRE...46..240F}. This is a ``forbidden" transition, with a very low transition probability for an individual atom. But the sheer abundance of \hi causes this signature to be observable. Due to the expansion of the Universe, the frequency (wavelength) of the radiation gets redshifted to lower frequencies (longer wavelengths)\footnote{The expansion of the Universe redshifts the frequency as $\nu_{obs}=\frac{\nu_{em}}{(1+z)}$, where $\nu_{obs}$ is the observed frequency for a radiation of frequency $\nu_{em}$ emitted at redshift $z$.}. It makes the signal observable at frequencies below $\lesssim$150 MHz, using radio telescopes. Radio telescopes can detect either the all-sky averaged ``global" signal using single dish total power instruments or detect the fluctuations in the signal using radio interferometers. 

The most sensitive low-frequency telescope for pursuing radio astronomy, Square Kilometer Array (SKA), was initially conceptualized to meet the required collecting area for studying the Universe using hydrogen \citep{Braun:2015B3}. However, in the subsequent years, the SKA has become one of the ``cornerstone facilities of the 21st century" to cater to several science goals (listed further on). Its unprecedented sensitivity and collecting area will address a wide range of questions in astrophysics and cosmology. The SKA Organisation has recently started construction of the telescopes at its two sites - Australia and South Africa \footnote{The readers are referred to \url{https://www.skatelescope.org/wp-content/uploads/2021/02/22380_SKA_Project-Summary_v4_single-pages.pdf} for an executive summary of the SKA Phase 1}. The SKA Low will be situated at the Murchison Radio-astronomy Observatory in Australia. It will be a low-frequency array operational between 50-350 MHz. The observatory will have 512 individual ``stations" along three spiral arms stretching up to 65\,km, with 256 dipole antennas in each station. On the other hand, SKA Mid (being built in the Karoo desert in South Africa) will operate with four different bands - 0.35-1.05 GHz, 0.95-1.76 GHz, 4.60-8.50\,GHz, and 8.30-15.30\,GHz. It will consist of an array of reflector antennas extending out to 150 km in a 3-arm spiral. There will be 133 dishes of 15m diameter and include an additional 64 antennas of 13.5m from the MeerKAT telescope. 

The combination of SKA-Low and Mid is expected to be the most sensitive radio telescope to shed light on several unanswered questions of the Universe. The key science drivers of the SKA include probing the CD and Epoch of Reionization EoR \citep{Koopmans:2015K0}, the study of dark energy and cosmology \citep{Maartens:2015Nk}, understanding fundamental physics with pulsars \citep{Kramer:20158P}, probing the transient Universe \citep{Fender:2015DT}, studying the formation and evolution of galaxies \citep{Prandoni:2015oN, Staveley-Smith:20151o}, cosmic magnetism \citep{Johnston-Hollitt:2015Cb}, etc. SKA-Low will probe the CD/EoR by studying the 21-cm signal from neutral hydrogen (\hi) at very high redshifts ($z\gtrsim6$). SKA-low will detect this signal statistically and image the \hi distribution at scales between arc-minutes to degrees for $z$ up to 28 \citep{Koopmans:2015K0}. Our understanding of cosmology, in general, will also be revolutionized using planned all-sky \hi intensity mapping for detecting total \hi emission from galaxies and all-sky continuum surveys for detecting galaxies to very high redshifts \citep{Maartens:2015Nk}. SKA will also provide deeper insight into galaxy formation and evolution through the star formation history, astrophysics of star formation and accretion processes, and studying the effect of black holes in galaxy formation \citep{Prandoni:2015oN}. Besides studying large-scale structures, the SKA will also be indispensable for studying small-scale fundamental physics and extreme gravity using pulsars \citep{Kramer:20158P}. Transient sources in the radio sky provide some of the best observational signatures of the most extreme phenomena in the Universe, like compact object mergers, stellar explosions, ultra-relativistic flows, to name a few. The unparalleled sensitivity of the SKA will observe these transient events to provide extremely sensitive probes into the underlying processes driving these events \citep{Fender:2015DT}. Besides these science goals, SKA will significantly impact other observational sciences like solar-terrestrial physics and planetary sciences. It will be an excellent instrument to work in synergy with other telescopes spanning a range of wavelengths in the electromagnetic spectrum to provide a very comprehensive picture of the processes driving the formation and evolution of the Universe at different scales.

The focus of this review is to provide an overview of the status of 21-cm experiments and determine the importance for development of an end-to-end simulation pipeline in the SKA era. The most sensitive operational and upcoming interferometers currently operationally and some under deployment have the observation of the redshifted 21-cm signal from CD/EoR as a major science goal. The sensitivity predictions indicate that they will have sufficient signal-to-noise to achieve statistical detection of the signal. However, tomographic imaging of the \hi in the IGM will be possible only with the SKA \citep{Koopmans:2015K0, Mellema:2015IS}. This review will mainly cover the challenges of observing this signal and demonstrate the usefulness of developing observational simulation pipelines to analyze actual array performance.

The review is organized in the following manner: in Section \ref{observation}, the basic physics of the 21-cm signal has been discussed. Section \ref{observation_global} describes the requirements, challenges and the current state of the art for 21-cm observations for global signal observations. The basic principles of statistical detection of the signal using interferometers is described in Section \ref{observation_ps}. The corresponding challenges are discussed in Section \ref{challenges}, and the recent upper limits placed on the 21-cm power spectrum are discussed in Section \ref{upper_interferometer}. Bearing in mind the challenges of statistically detecting the 21-cm signal using interferometers, Section \ref{sims} describes the end-to-end simulation pipeline developed for studying realistic observational effects. The performance of the pipeline and its limitations is demonstrated in Section \ref{perf_pipe}. Finally, he paper is concluded with a summary in Section \ref{summary}.

\section{Physics of the Observable Signal}

\label{observation}

The 21-cm line is a hyperfine transition in atomic (neutral) hydrogen. The aligned electron-proton spin state is a higher energy state than the anti-aligned one. The energy difference between the two states is 5.87 $\times$10$^{-6}$eV, corresponding to a photon of wavelength $\sim$21-cm. The transition of these states is mediated by the absorption or emission of a photon of the same energy.

The spin temperature $T_{s}$ represents the effective strength of the 21-cm line. It is the effective excitation temperature for the hyperfine transition and determines the emission and absorption properties of the 21-cm line. It is defined in terms of the Boltzmann factor for the relative occupancy of the triplet with respect to the singlet state for the ground state of hydrogen atom as \citet{Field1958PIRE...46..240F}:
\begin{equation}
    \frac{n_1}{n_0} = 3 \exp\big\{\frac{-T_*}{T_s}\big\}
\end{equation}
where $n_0$ and $n_1$ are the occupancy of the singlet and triplet states and $T_*$=0.0682 K. 

$T_{s}$ is governed by an underlying series of complex physical processes (see comprehensive reviews by \citet{Furlanetto2006} \& \citet{Pritchard2012} for detailed discussions on the different factors controlling $T_{s}$). The observation of the cosmological signal is done through its contrast against the CMB -  $T_{s} >$ $T_{\gamma}$ gives an emission signature and $T_{s}<T_{\gamma}$ gives an absorption signature wherein $T_{\gamma}$ is the CMB temperature.

The fundamental observable quantity is the differential 21-cm brightness temperature $\delta T_b$. From the radiative transfer equation, $\delta T_b$ is the temperature for an equivalent black-body radiator. For \hi 21-cm line, $\delta T_b$ is given by : 

\begin{equation}
 \delta T_{b}(\nu) = T_s(1-e^{-\tau_\nu}) + T_{\gamma}(\nu)e^{-\tau_\nu}
    \label{soln_rad_transfer}
\end{equation}

where $\tau_{\nu}$ is the optical depth of the 21-cm radiation at redshift $z$, given by:

\begin{equation}
    \tau(z) = \frac{3c\lambda_{21}^{2}h_{p}A_{10}n_{HI}}{32\pi k_{B}T_{s}(1+z)(\partial v_{r}/\partial r)}
\end{equation}
where $h_p$ is the Planck's constant, $A_{10} = 2.85\times 10^{-15} s^{-1}$ is the spontaneous decay rate of the hyperfine transition, $n_{HI}$ is the number density of hydrogen atoms and $\partial v_{r}/\partial r$ is the line-of sight radial velocity, where $v_{r}$ is the physical radial velocity and $r$ is the comoving distance. In a completely neutral and homogeneous Universe, $n_{HI}=\Bar{n}_{H}(z)$ and $\partial v_{r}/\partial r = H(z)/(1+z)$ , where, $H$ is the Hubble parameter. 

Plugging in the value of $\tau_{\nu}$ in Equation \ref{soln_rad_transfer}, the differential brightness temperature is:

\begin{align}
  \delta T_{b}(\nu) & =  \frac{(T_s-T_{CMB})}{(1+z)}e^{-\tau} \nonumber
  \approx  \frac{(T_s-T_{CMB})}{(1+z)}\tau \\
   & \approx 26.8 \mathrm{mK} \Big(\frac{\Omega_{b}h}{0.0327}\Big) \Big(\frac{\Omega_{m}}{0.307}\Big)^{-1/2} \Big(\frac{1+z}{10}\Big)^{1/2} \nonumber \\ & \hspace{120pt} \Big(\frac{T_{s}-T_{CMB}}{T_s}\Big)
   \label{deltb}
\end{align}

\begin{figure*}
    \includegraphics[width=16cm,height=8.5cm]{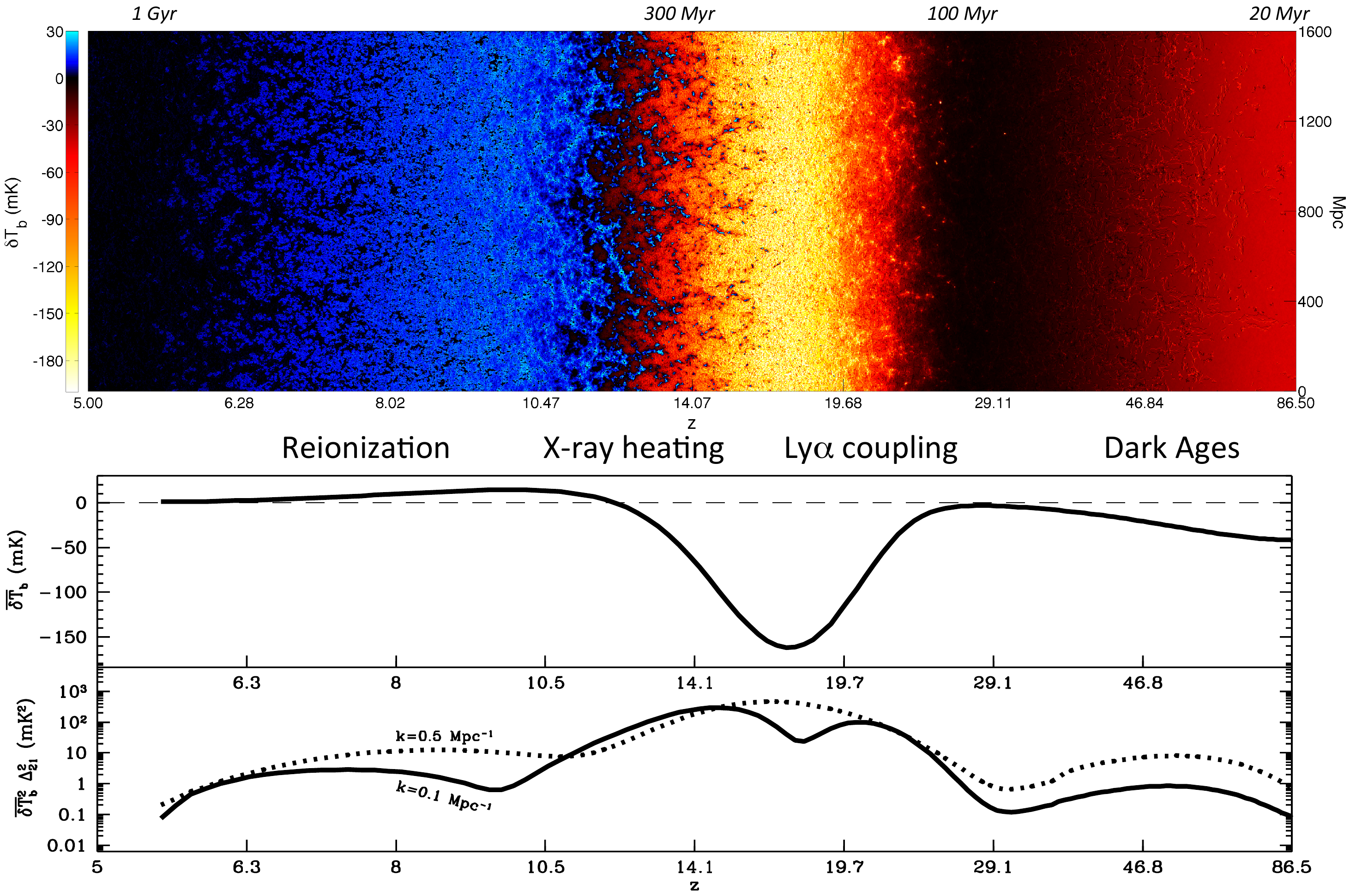}
    \caption{The evolution of the 21-cm signal in the redshift range $90\lesssim z\lesssim7$. The top panel shows a slice through a  21-cm brightness temperature cube, displayed in contrast with the CMB. The middle panel shows the predicted global signal of the 21 cm line. The botom panel shows the spherically averaged PS evolving as a function of redshift for two spatial wavenumbers, $k$ = 0.1 Mpc$^{-1}$ and 0.5 Mpc$^{-1}$. The Figures are generated from the 21cmFAST simulations \citep{21cmf} using a fiducial model of ionizing sources. The images are available publicly at \url{http://homepage.sns.it/mesinger/EOS.html}. }
    \label{demo}
\end{figure*}

Observations of the 21-cm signal from the early Universe using $T_{b}$ can be done using two approaches. One is via measurement of the global signal, where $T_{b}$ is measured by averaging over the entire sky. The other is via measurements of the fluctuations in the full $T_{b}$ field, either quantifying the statistical properties (power spectrum) or reconstruction of the signal in three dimensions (tomography). Figure \ref{demo} demonstrates the time evolution of the 21-cm signal under some fiducial assumptions of the nature of ionizing sources and the evolution of the physical parameters controlling the signal generated from the semi-numerical signal generation package 21cmFAST \citep{21cmf}. The top panel is the slice through a three-dimensional signal cube, the middle panel depicts the global signal evolution, and the bottom panel shows the power spectrum (PS) evolution with redshift. In the following subsections, we provide an overview of the aforementioned observational techniques and how they are used for detecting the cosmological 21-cm signal. 

\section{Precision Radio Cosmology Observations : Global Signal}
\label{observation_global}

Global signal measurements are typically made with sensitive single element radiometers, although there exist interferometric techniques to study this all-sky averaged signal \citep{ZEBRA, SS2015, 2015ApJ...809...18P}. There exist mature ground-based experiments seeking a first detection of the global signal from CD/EoR including SARAS \citep{SARAS3}, EDGES \citep{EDGES2018}, LEDA \citep{Bernardi2018}, CTP \citep{Nhan2019}, and REACH \citep{REACH2019}. Most of these experiments observe the sky with a large beam, letting the sky drift across the beam, recording spectra at regular intervals. Due to the all-sky nature of the signal, the collecting area of the radiometer is not the limiting factor. Instead, the emphasis here is on achieving a precisely calibrated radio sky-spectrum at a location with minimal radio-frequency interference (RFI). Most single radiometer experiments comprise three sub-systems: an antenna with a wide beam, an analog receiver with calibration electronics, and a digital receiver that digitizes the data and generates a channelized spectrum. The details of the implementation, operating bandwidths, and calibration strategies vary across experiments. All experiments have a common goal of calibrating the receiver-induced bandpass to accuracy levels that do not introduce spurious artifacts in the measured data set. In post-observation data-processing, the foregrounds may be cleanly separated from the CD/EoR signal. This is a tall task since the dynamic range of observations can range from $10^4$ to $10^6$, depending on the region of the sky being observed and the strength of the signal. 

\subsection{Foregrounds for Global 21-cm Experiment}
The primary challenges in global signal detection may broadly be classified into astrophysical and instrument-based. Astrophysical challenges arise from separating foregrounds from the cosmological signal, the foregrounds being dominated by Galactic synchrotron emission. While a power law can model the synchrotron emission from point sources and localized regions in the sky, the summation of power laws as observed by a large beam (such as in the case of global EoR detection experiments) can no longer be modeled as a power law. However, over sufficiently wide bandwidths, the global signal from CD/EoR has multiple turning points absent in the average foreground spectrum, which is spectrally `smooth'. This distinctive feature unique to the cosmological signal can separate them from foregrounds that are $4-6$ orders of magnitude brighter. The separation of the CD/EoR signal from foregrounds poses two distinct challenges. First is the requirement to measure foregrounds to the accuracy required to model them independently. To estimate the expected template of foregrounds, sky models typically use existing all-sky radio maps spaced in frequency. While sky-models such the Global Sky Model \citep{GSM} and the improved Global Sky Model \cite{IGSM} use data-driven methods to interpolate the spectrum between the frequencies, GMOSS \citep{GMOSS} uses a physically motivated model based on radiative processes instead. The key to an effective sky model is one that does not introduce any artifacts by way of unphysical spectral features in the spectrum simulated using the model. Interestingly, global CD/EoR experiments designed to be sufficiently sensitive to detect the cosmological signal are best suited to improve foreground models. These have come in the form of (re-)calibration of all-sky maps \citep{monsalve2021,patra2015} and improved measurements to spectral indices of emission towards different directions \citep{mozdzen2019,rogers2008}. With an expectation of foreground emission spectrum, the next challenge is the separation of this contaminant bright signal from the faint signal from CD/EoR. Traditionally foregrounds have been modeled as low-order polynomials; however, this stands the risk of `over-fitting'. Increasing the order of polynomials to fit the data better stands the risk of subsuming more of the CD/EoR signal, in addition to introducing additional turning points in the resulting residual on (polynomial-based) foreground subtraction. To alleviate this problem, Maximally Smooth (MS) functions \citep{MSR2015} have been proposed to model foregrounds as they minimally subsume the CD/EoR signal and preserve the turning points, which are critical to understanding the science of CD/EoR. It has been demonstrated using GMOSS simulations that the foregrounds are well described by MS functions, whereas the CD/EoR signal is not \citep{MSR2017b}. Thus, MS functions or variants therein can be used to separate foregrounds from sky-spectra.

\subsection{Instrumental Systematics}
Instrument-based challenges in global CD/EoR signal detection can arise from systematics or artifacts in the experiment that confuse signal detection. For instance, an experiment employing a frequency-dependent antenna beam can introduce `mode-coupling', which is the mixing of spatial features with spectral features. A beam that looks at different regions of the sky (due to chromaticity) at different frequencies will naturally have different total measured power as a function of frequency. The resultant spectral structure in the observed sky-spectrum is an artifact introduced by virtue of the antenna properties and not inherent to astrophysical processes of the emission. Yet another source of confusing artifacts is impedance mismatch induced standing wave structures in the spectrum. These, being `additive' in nature, are not readily calibrated in most noise-injection-based calibration schemes. A few more noteworthy challenges facing CD/EoR detection experiments are (a) terrestrial radio frequency interference (RFI), (b) refraction, emission, and absorption effects of the ionosphere, and (c) the effect of objects in the near-field of the antenna, such as the ground beneath, on the antenna properties.

\subsection{Radio Frequency Inteference}
Global CD/EoR experiments take great care in observing at the most radio-quiet locations with minimal RFI. However, contamination from RFI is still observed in the form of low-lying RFI that appears after several hours of integration or scattering into the antenna beam from outside the line-of-sight or downlink transmission from satellites when they pass overhead. Localized RFI in data can result in channel loss by means of flagging or loss of entire spectra when the RFI is broadband or extremely bright. Confusing artifacts in recorded sky-spectra can also result from the `coupling' of objects in the near-field of the antenna to antenna properties. Most significantly, edge effects of metallic ground planes beneath the antenna or the stratified and non-homogeneous soil over which the antenna observes can result in spurious spectral structure. This has necessitated novel solutions. For instance, the SARAS3 observes on the water of a sufficiently deep freshwater lake of suitable conductivity.

\subsection{Ongoing Global Signal Experiments}
The most conducive environment for observing the global CD/EoR signal is one devoid of terrestrial RFI, low-to-no ionospheric effects, and a uniform medium. The lunar farside provides such an environment. The Moon is expected to attenuate terrestrial RFI, providing a radio-quiet environment in the shadow region of the Earth and Sun \citep{WINDWAVES}. Several experiments have been proposed or planned to operate in the lunar farside, taking advantage of the pristine environs. These include Dark Ages Polarimeter PathfindER (DAPPER) \citep{DAPPER}, Lunar Surface Electromagnetics Experiment (LuSEE) \citep{LUSEE2020}, Dark Ages Reionization Explorer (DARE) \citep{DARE}, the ,  Radio wave Observations at the Lunar Surface of the photoElectron Sheath (ROLSES) \citep{ROLSES} and FARSIDE \citep{FARSIDE}. PRATUSH -- Probing ReionizATion of the Universe using Signal from Hydrogen is a cosmology experiment from India proposed to detect the global 21-cm signal from CD/EoR in orbit around the moon. With a goal bandwidth of 40-200 MHz and a baseline bandwidth of 55-110 MHz, PRATUSH is currently in the pre-project studies phase, funded under the Indian Space Research Organisation in response to an announcement of opportunity for science payloads in 2018.

While there are several ongoing and proposed experiments for global CD/EoR signal detection, a conclusive high confidence detection remains at large. The faint nature of the signal, the complexity of realizing a truly frequency-independent antenna over wide bandwidths (greater than an octave), and the challenges posed by systematics make this a difficult task. Furthermore, the wide range of plausible signals predicted by standard models alone results in poor constraints on the exact nature of the signal being investigated, making the problem all the more challenging. The parameter space becomes even larger when including non-standard physics. The need of the hour is to have custom-designed instruments for precise, well-calibrated measurements of the sky-spectrum and multiple concurrent measurements that will enable obtaining a strong handle on systematics and RFI mitigation. Detecting the global signal from CD/EoR is currently outside the purview of the SKA, which focuses on the interferometric methods of power-spectrum measurement and, ultimately, tomography. However, a concept of detecting the global CD/EoR signal detection with the SKA is explored in this issue (Sathyanarayana Rao et al.). The significance of the global CD/EoR detection and the advantages of observing it with the SKA are explored therein. Additionally, the strategy most suited to such a measurement is also proposed. Detecting both the global signal along with the spatial fluctuations from CD/EoR with the SKA would truly be a resounding result coming from the mega-telescope! 

\subsection{Global Signal: Way Forward}

Despite the challenge at hand, global CD/EoR signal detection experiments have led the way in placing constraints on the astrophysics of the processes involved. For instance, EDGES has ruled out $tanh$ based reionization models of duration $\Delta z \lesssim 2$ \citep{monsalve2017}. SARAS has rejected 10\% of standard reionization models, most of these being late-heating rapid reionization models \citep{SARAS2}. In 2018 EDGES reported detecting a deep and wide absorption trough centered around 78 MHz, which if of cosmological origins does not conform to any standard models. This generated tremendous interest in the community, seeking a range of physical explanations for this signal, including excess background radiation and radiatively interacting dark matter, among others \citep{barkana2018,fialkov2019}. The result was questionable,  with the signal being attributed to non-cosmological origins, including those arising from artifacts of analysis methods, standing waves from cable-lengths internal to the system, and chromaticity from environmental factors, including the antenna ground plane \citep{singh2019,hills2018}. Recently, as of the writing of this paper, SARAS has experimentally rejected the non-standard detection from EDGES with confidence exceeding 95\% \citep{SARAS3}, determining that the reported signal was not of cosmological origin. With confidence in standard models restored, the task of detecting the true global signal from CD/EoR remains.

\section{Precision Radio Cosmology Observations : Power Spectrum}
\label{observation_ps}

The challenges for interferometric detection of the \hi 21-cm signal are not drastically different from those described above for the global signal. However, since radio interferometers observe fluctuations in the signal, the manifestation of the corruptions differs. In this subsection, we explore the 

Interferometers targeting the statistical detection of the 21-cm signal are designed keeping a few features in mind. They must be extremely sensitive in their design and the observing strategies to maximize the probability of detecting the faint signal. The instruments must also be sensitive to angular scales of arcminutes at $\nu \lesssim$150 MHz. The cosmological \hi signal can map the Universe in 3D since it evolves with redshift, hence frequency. Thus the instrument targeting this signal should have a sufficiently large bandwidth. The observations also require stable electronics, which minimizes the impact of systematics.

An interferometer observing the sky measures voltage fluctuations of the incoming electromagnetic wave. The Van Cittert-Zernike theorem expresses the relationship between the incoming radiation and the actual measured quantity as \citep{Thompson_book}:

\begin{equation}
     V ({\mathbf{U}},\nu) = \int \int A(\hat{\mathbf{s}},\nu) B(\nu) I(\hat{\mathbf{s}},\nu) e^{-i2\pi \nu \mathbf{U}.\hat{\mathbf{s}}} d\Omega,
     \label{vis_eq}
\end{equation}

\noindent where, ${\mathbf{U}}$ is the vector of the distance between two antennas, i.e. baseline, $I(\hat{\mathbf{s}},\nu)$ is the actual brightness distribution in the sky, $ A(\hat{\mathbf{s}},\nu)$ antenna beam pattern as a function of frequency ($\nu$) and  $B(\nu)$ is the instrumental bandpass response; the unit vector $\hat{\mathbf{s}} \equiv (l,m,n)$, where $l,m,n$ are the direction cosines towards east, north and zenith respectively with $n = \sqrt{1-l^{2}-m^{2}}$ and $ d\Omega = \frac{dl dm}{\sqrt{1-l^{2}-m^{2}}}$.

\begin{figure*}
    \includegraphics[width=18cm]{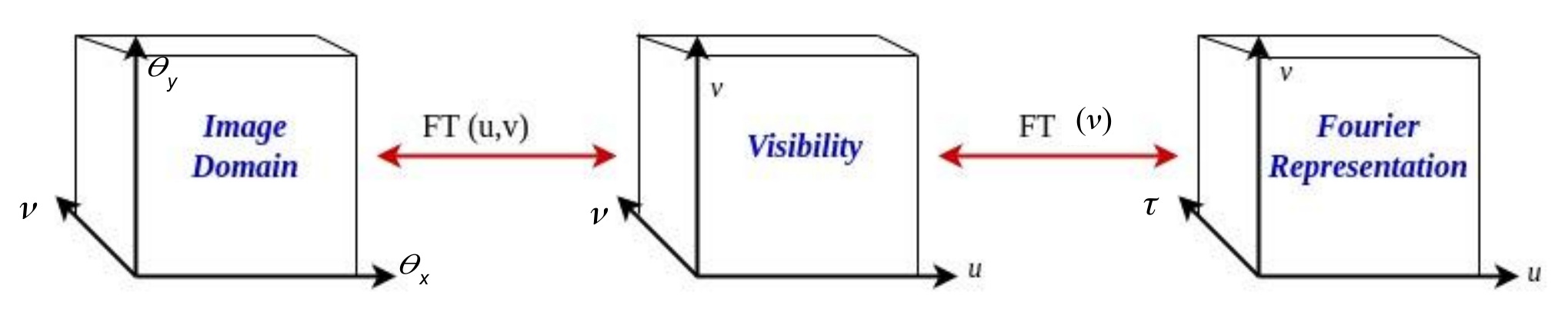}
    \caption{Fourier conjugate relationships between the fundamental observable of an interferometer - visibility with image (Fourier transform along the plane of sky) and the Fourier representation (Fourier transform along frequency direction). The image cube allows for pinpointing source location and Fourier representation allows for analysis of spatial structure of the signal.}
    \label{ft_ps}
\end{figure*}

The quantity $V ({\mathbf{U}},\nu)$ in the left-hand side of Equation \ref{vis_eq} is the ``visibility", the quantity measured by an interferometer, which is the Fourier transform of the intensity distribution in the sky. Thus, performing an inverse Fourier transform should provide the sky intensity. The inverse transform can be done along both frequency ($\nu$) and spatial ($\mathbf{U}$) directions. This is depicted in Figure \ref{ft_ps}. Statistical detection of the signal through its power spectrum can be obtained both in the image domain and the Fourier domain. Inverse Fourier transform of $V ({\mathbf{U}},\nu)$ along frequency delay domain visibility, $V ({\mathbf{U}},\tau)$. Using this formalism, the cylindrical PS, with unit $\mathrm{K^2(Mpc/h)^3}$ (as discussed in \citealt{Morales2004}) is given by:

\begin{equation}
    P(\mathbf{k}_{\perp}, k_{\parallel}) = \Big(\frac{\lambda^{2}}{2k_{B}}\Big)^{2}    \Big(\frac{X^{2}Y}{\Omega B}\Big)  |V (\mathbf{U},\tau)|^{2}, 
    \label{2dps_eq}
\end{equation}

\noindent where $\lambda$ is the wavelength of observation, $k_{B}$ is the Boltzmann constant, $\Omega$ is the primary beam response, B is the observing bandwidth, X and Y are the conversion
factors from angular and and frequency directions to the transverse co-moving distance (D(z)) and the co-moving depth along the line of sight, respectively \citep{Morales2004}. In Equation \ref{2dps_eq},  $\mathbf{k}_{\perp}$ and $k_{\parallel}$ are the Fourier modes (or spatial wave numbers or k modes) perpendicular and parallel to the line of sight, given by: 
 \begin{equation}
 \mathbf{k}_{\perp} = \frac{2\pi |\mathbf{U}|}{D(z)} \hspace{20pt} \& \hspace{20pt}
  k_{\parallel} = \frac{2\pi \tau \nu_{21} H_{0} E(z)}{c(1+z)^{2}}
  \label{k_eq}
 \end{equation}
 
\noindent where $\nu_{21}$ is the rest-frame frequency \hi 21-cm transition (i.e. 1400 MHz), $z$ is the redshift corresponding to the observing frequency, $H_{0}$ is the Hubble parameter and $E(z) \equiv [\Omega_{\mathrm{M}}(1+\textit{z})^{3} +  \Omega_{\Lambda}]^{1/2}$, with $\Omega_{\mathrm{M}}$ and $\Omega_{\Lambda}$ being the matter and dark energy densities, respectively \citep{Hogg1999}. 

\begin{figure}
    \includegraphics[width=\columnwidth]{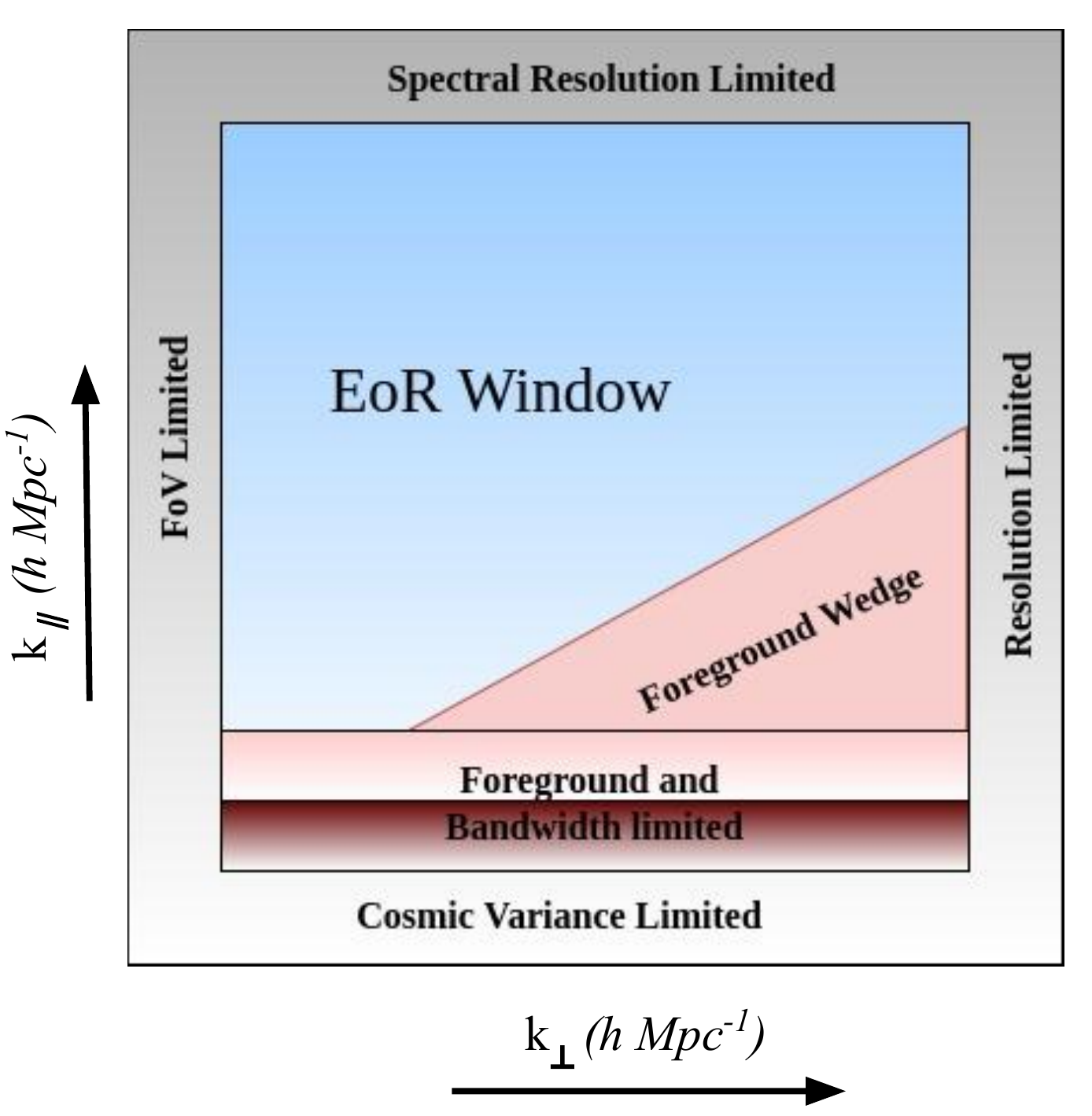}
    \caption{Schematic diagram for the Fourier space accessible by a radio interferometer, along with the factors limiting accessible modes.}
    \label{wedge_window}
\end{figure}

From Equation \ref{k_eq}, it can be seen that the accessible k modes depend on the interferometer design. The angular wave numbers, $\mathbf{k}_{\perp}$, are related to the instrument layout, with the accessible modes are determined by the baseline distribution. The longest baseline limits the finest accessible angular scales (largest $k_{\perp}=|\mathbf{k}_{\perp}|$), while the shortest ones limit the largest accessible $k_{\perp}$ value. The lines of sight wave numbers, $k_{\parallel}$, are determined by the spectral features of the instruments. The finest modes (with highest $k_{\parallel}$) are spectral resolution limited, and coarsest modes are bandwidth limited. These limits are depicted qualitatively in Figure \ref{wedge_window}. For 21-cm experiments, the accessible region in the $\mathbf{k}_{\perp} - k_{\parallel}$ plane is called the ``EoR window" (blue region in Figure \ref{wedge_window}). The EoR window is expected to be free from contamination by the different corruption sources (discussed in detail in the next section). However, there is still noise in the EoR window, which is reduced only by averaging over long observing times. The contamination due to foreground power is the strongest at the lowest $k_{\perp}$ modes and decreases towards higher $k_{\perp}$ modes. This lines of constant contamination in the Fourier plane is given by $\mathbf{k}_{\perp} \propto k_{\parallel}$, giving the characteristic ``wedge" shape (light orange region in Figure \ref{wedge_window}) \citep{Datta2010, Morales2012}. This region arises due to the inherently chromatic response of the interferometer, and while the overall shape remains the same, the profile details are complicated by the presence of different primary beam and foreground effects. Specifically, corruptions due to incorrect foreground and instrument modeling or calibration errors cause the foreground power to be leak into modes outside the wedge, an effect called ``mode" mixing \citep{Morales2012}. 

The presence of the foreground wedge gives rise to a strategy for detecting the \hi 21-cm power spectrum by taking measurements outside the wedge and in the EoR window only. This is called the ``foreground avoidance" method. The avoidance method is apparently easier than the subtraction method (where foregrounds are modeled in great detail to remove them from the data). However, the orders of magnitude brighter foregrounds leaking into the wedge by even a tiny amount can render detection difficult due to mode-mixing. The other issue with avoidance is, of course, giving up certain k-mode, especially at low k values. This results in loss of sensitivity \citep{chapman2016}. Currently, hybrid approaches using both avoidance and subtraction are being explored \citep{Kerrigan_2018}. 

The cylindrical PS is averaged in spherical bins, either using all the $\mathbf{k}_{\perp}$ modes or using only the k-modes outside the wedge. From Equation \ref{2dps_eq},  $P(\mathbf{k}_{\perp}, k_{\parallel})$ can be averaged spherically averaged in independent $k$-bins to produce the 1D power spectrum given by \citep{Morales2004} :

\begin{equation}
    \Delta^2(k) = \frac{k^3}{2\pi^2} \langle P(\bf k)\rangle_{k}
    \label{3d_ps}
\end{equation}

\noindent where $k = \sqrt{\mathbf{k}_{\perp} + k_{\parallel}^{2}}$. The 1D PS is the ultimate target of interferometric observations. The shape and amplitude of the spherical PS are controlled by the different physical parameters that control the evolution of the signal over time. Hence, it is used to constrain the different astrophysical parameters. Errors in PS estimation thus lead to the wrong estimation of parameters and constraints on different models of reionization erroneously. 

In the following subsection, the challenges with handling the different sources of error while performing sensitive radio observations.

\section{Observational Challenges for Interferometric 21-cm Observations}
\label{challenges}
Interferometers are complex instruments. The extraterrestrial radio signals received by an interferometer combine with the instrumental systematic effects, making it challenging to detect the target signal. In the case of the cosmological \hi signal, it is further complicated by the orders of magnitude brighter astrophysical foregrounds. While we have a good grasp of the nature of foreground emissions and how they vary with frequency and time, the knowledge is not perfect. Similarly, for instrumental properties, certain things we understand correctly and certain others are undetermined. The known systematics can, in principle, be dealt with during data reduction, though this is often computationally challenging. Additionally, ``optimally" dealing with systematics can lead to sensitivity loss and have other unwanted effects on the data. The unknown systematics are even more harmful, as they might lead to uncorrected mode proliferation, biasing the analysis. The 21-cm observations thus need analysis strategies that can deal with the known contaminants and minimize the unknown systematics to get unbiased results. 

There have been multiple works in the past decade that have demonstrated explicitly how the presence of different systematic errors and mismodelled terms affect the recovery of EoR data sets, as well as methods for their mitigation for signal recovery. These include studies on foregrounds \citep{Datta2009, Datta2010, trott2012, Vedantham_2012, chapman2016, Thyagarajan_2015a, Thyagarajan_2015b, mertens2018, hothi2020}, improving calibration model \citep{offringa_2015, Barry2016,trottwayth2016,Patil2016, procopio_2017, Ewall2017, dillon2018,orosoz, Kern_2019}, instrumental model and systematics \citep{Datta2009, Datta2010, thayagrajan2013, eloy2017, trott2017, Li_2018}, and ionospheric effects \citep{vedantham_iono, jordan, Trott_2018}. 

\subsection{Foregrounds}
One of the most potent contaminants of the redshifted 21-cm signal from the CD/EoR is the astrophysical foregrounds. They constitute all the emissions in the radio sky brighter than the target signal present at the observing frequency. They include diffuse galactic synchrotron emission (DGSE), galactic and extragalactic free-free emission, and emission from astrophysical sources, viz. star-forming galaxies (SFGs) and active galactic nuclei (AGNs). These emission sources themselves are of significant scientific interest (for example \citealt{Haslam1, Haslam, costa2008, guzman, remazeilles, dowell2017,zheng2018, meerklass, southern_dgse, Intema16, lotss_dr1, gleam, lobes}) and are currently being studied actively with state of the art radio telescopes. However, for 21-cm cosmology, these contaminants need to be removed from the relevant data sets. DGSE is the dominant emission at frequencies $\sim$ 150 MHz at angular scales $\gtrsim$degree, while the extragalactic sources dominate at smaller angular scales. Since radio emissions arise from synchrotron processes, it is expected that these sources are spectrally smooth, i.e., do not have frequency structures contrary to the \hi signal \citep{Shaver1999}.

\begin{figure}
    \includegraphics[width=\columnwidth, height=6cm]{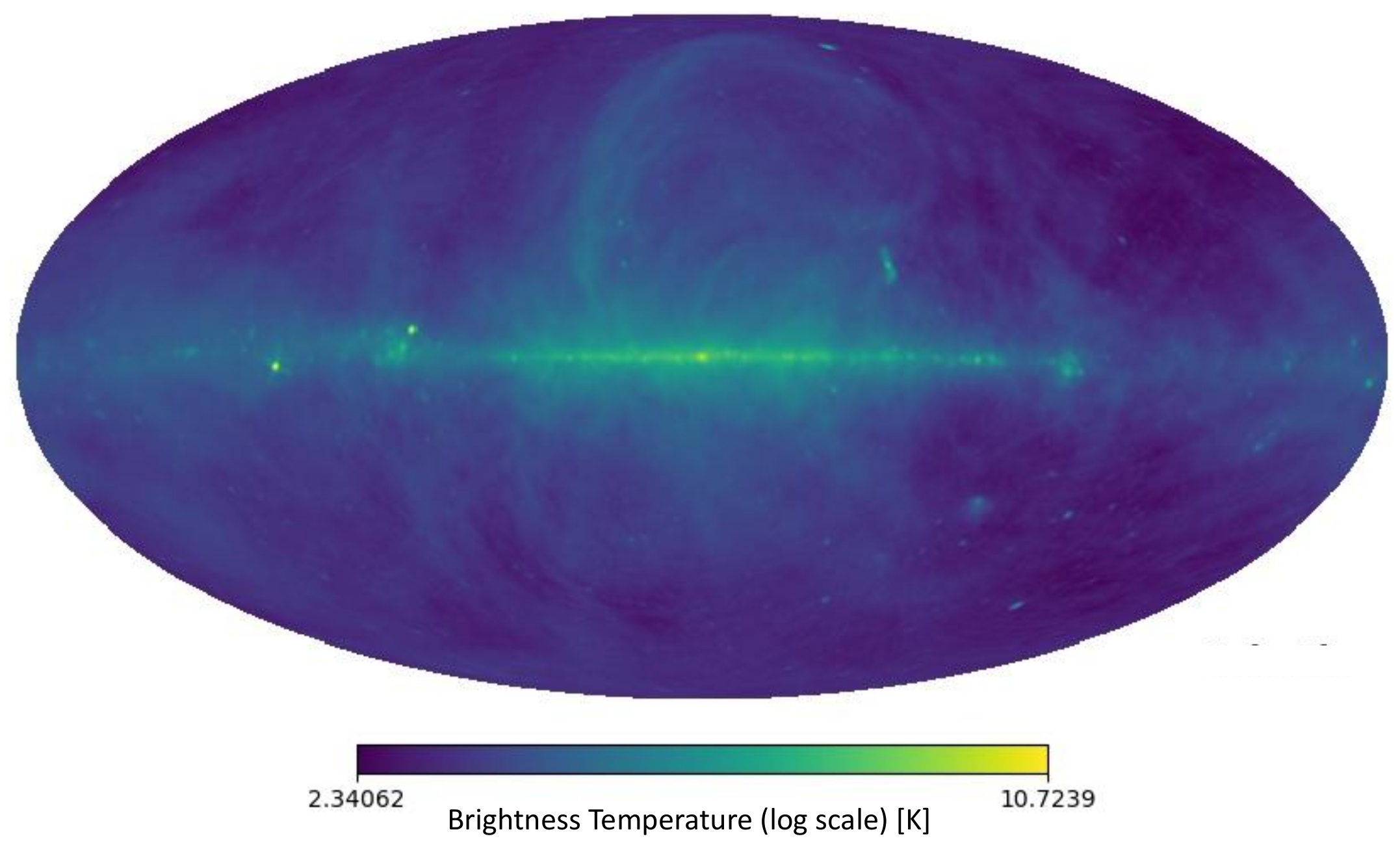}
    \caption{408 MHz Haslam map \citep{Haslam1, Haslam, remazeilles} of DGSE. The emission is clearly bright at the galactic plane, and is brighter at lower frequency. The color bar is in log$_{10}$ scale for temperature in K and is brighter than the 21-cm emission. The plot is made using the publicly available PyGDSM package (\url{https://github.com/telegraphic/pygdsm}).}
    \label{fg}
\end{figure}

The major portion of the foreground emission at angular scales $\gtrsim$degree for $\nu \lesssim$ 150 MHz consists of DGSE. Figure \ref{fg} shows the brightness temperature map of the DGSE at 408 MHz made by \citet{Haslam1, Haslam}. It is seen that the brightness temperature of this emission (plotted as the logarithm of temperatures in K) is brightest at the galactic plane. However, there are observations of excess synchrotron power on angular scales $\sim$ degrees at ``colder" regions near the North Celestial Pole and South Galactic poles \citep{Bernardi2010, Lenc2016, lofar_l1}. The average brightness temperature is between a few 10s to 100s K. The frequency dependence of the DGSE brightness temperature is approximated as a power law of the form $T(\nu)\propto\nu^{-\alpha}$. Thus, the temperature would be much higher at lower frequencies. Besides synchrotron radiation, free-free emission and a small amount of radio emission from radio haloes and relics also contribute to the diffuse foreground to some extent. 

\begin{figure*}
    \includegraphics[ width=15cm,height=8.5cm]{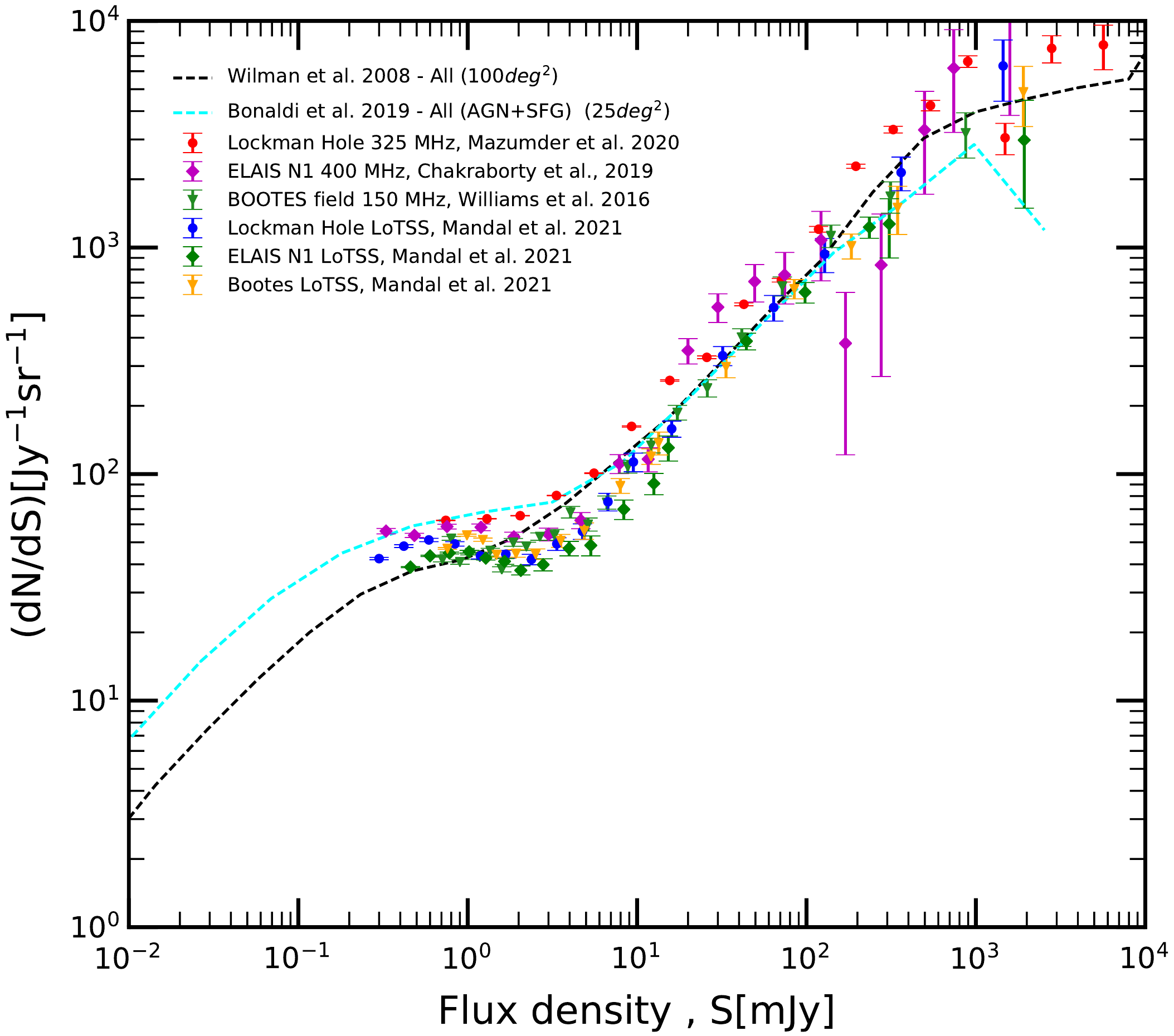}
    \caption{Euclidean normalised differential source counts at 150 GHz from LoTSS survey for Lockman Hole (blue circles), ELAIS-N1 (green diamonds) and Bo\"{o}tes (orange inverted triangle) fields from \citet{Mandal2021}. Source counts from other observations of the same field - Bo\"{o}tes (green inverted triangles, \citealt{WilliamsBootes}) at 150 MHz from LOFAR, Lockman Hole (red circles, \citealt{aishrila1}) at 325 MHz from GMRT and ELAIS-N1 (magenta diamonds, \citealt{arnab2}) at 400 MHz from from GMRT. Comparison with source counts from simulations - SKADS (black dashed curve, \citealt{S3}) and T-RECS (cyan dashed curve, \citealt{trecs}) has also been done.}
    \label{compact}
\end{figure*}

Extragalactic foregrounds are mostly compact sources consisting chiefly of AGNs and SFGs. Similar to the diffuse foregrounds, the extragalactic radio galaxy populations themselves are of immense scientific interest. The low-frequency population of radio sources is not yet adequately constrained, especially at the faint flux density end. There have been many surveys that have tried to provide a consensus on the number counts, luminosity functions, and other properties \citep{padovani2015, Prandoni2018, HARDCASTLE2020101539}. The recent results from the LOFAR Two-Meter Sky Survey (LoTSS) have also started to explore the counts at the 150 MHz low flux end(see \citet{Mandal2021} and references therein). The current deepest 150 MHz source counts from three extragalactic deep fields - Lockman Hole, ELAIS-N1, and Bo\"{o}tes, obtained using the LoTSSS data \citep{Mandal2021} is shown in Figure \ref{compact}. They are overlaid with counts from other surveys- Bo\"{o}tes at 150 MHz using LOFAR \cite{WilliamsBootes} and GMRT observations of the ELAIS-N1 at 400 MHz \citep{arnab2} and Lockman Hole at 325 MHz \citep{aishrila1, am_clustering}. The AGNs dominate the flux densities $\geq$1 mJy while SFGs dominate at the sub-mJy level. This is evident from the signature flattening of the counts at sub-mJy flux density levels. In addition to source counts, it is also essential to understand their spatial distribution (i.e., clustering). Ignoring clustering may result in underestimating the foreground power, leading to confusion in detecting the cosmological signal. It has been shown by \citet{dimatteo2004} that spatial clustering of extragalactic sources dominates the fluctuations for angular scales $\theta \gtrsim 1'$ at 150 MHz (for flux density $\gtrsim$0.1 mJy). Radio spectra of sources also need to be studied and modeled well since any deviation from the predicted smooth nature will make their removal challenging. Thus, a detailed study of the compact extragalactic sources in terms of the spatial, flux, and frequency characteristics will be required for EoR data sets.

The foreground contamination is generally removed by considering them to be ``spectrally smooth", i.e., their spectrum is a smooth function of frequency (see for example \citet{Tegmark_2000, Planck2016}). There are different foreground mitigation approaches that employ the spectral smoothness assumption. These algorithms use parametrized or non-parametrized strategies to model out foregrounds. Some approaches include principal component analysis, generalized morphological component analysis, independent component analysis, Gaussian process regression, etc. \citep{gmca, fastica, mertens2018}. Each process comes with its pros and cons. Specifically, a general caveat needs to be considered for foreground removal - the presence of non-smooth foreground components in the residuals, which hamper the signal detection. Even for truly smooth foregrounds, this may be caused by the instrumental feature introducing some non-smooth features. There are also chances of signal removal along with foregrounds. The foreground contamination can also be dealt with using avoidance (discussed in detail in Section \ref{observation_ps}). Thus, handling foregrounds poses a major challenge for sensitive radio observations targeting the \hi 21-cm signal from CD/EoR.

\subsection{Systematics}
Another vital source of error comes from calibration of the low-frequency data. Meticulous research has shown that minute errors in instrumental calibration affect the separation of signals from the bright foreground emission. The basic equation for calibration (assuming implicit time and frequency dependence) is thus given by \citep{Taylor1999}:
\begin{equation}
    {V}_{ij}^{obs} = g_{i}g_{j}^{*}V_{ij}^{true} + n_{ij}
    \label{gains_eq}
\end{equation}
\noindent where ${V}_{ij}^{obs}$ and $V_{ij}^{true}$ are the observed and actual visibilities at time t, $g_{i}, g_{j}^{*}$ are the antenna-based complex gains ($g_{j}^{*}$ is the complex conjugate of $g_{j}$) at time t, and $n_{ij}$ is the noise; the subscripts i and j represent antenna pairs in the array. 

The gain term should be unity in an ideal scenario, but that is not possible with real instruments. Since we want to determine$V_{ij}^{true}$, a solution for the gain is required. Calibration solves for these complex instrumental gains to derive the true visibilities. There are N(N-1)/2 baselines, and hence that many independent measured visibilities for an array with N antennas. We need solution for all the N(N-1)/2 true visibilities along with N gain  factors ($g_i$ and $g_j$ in Equation \ref{gains_eq}). This entails making assumptions about the sky as well as the instrument. 

The traditional calibration approach in radio astronomy is the ``sky-based" calibration approach. In this method, a sky model is used as a prior and simulated through the instrument model to obtain a set of visibility models. This model becomes equivalent to the ``true" visibility, iteratively solving for the gains until the solutions converge. However, the sky model constructed for sky-based calibration may have mismodelled or missing sources, resulting in imperfect calibration solutions. These imperfections affect the observational data sets targeting the EoR and make signal detection difficult (see, for example, \citealt{Barry2016, Patil2016, Ewall2017}). There is another calibration approach being explored for calibrating interferometric data for sensitive observations. This is known as the redundant calibration approach \cite{Wieringa1992, Liu2010}, and works for arrays like HERA, which have highly regular layouts with many redundant baselines. The redundant calibration method adopts a prior that the actual visibility of the redundant baselines is equal. Thus instead of modeling the true sky visibilities, this method solves for both them along with the gains. Since arrays with highly regular layouts have more measured visibilities than unique baselines, the system is overdetermined despite leaving the true visibility as free parameters. However, the redundant approach works if and only if the array layout is highly regular and has perfect redundancy with identical primary beams. In the absence of perfect redundancy, calibration using this approach can also adversely affect data analysis for sensitive observations.

From Equation \ref{vis_eq}, we can see that in order to obtain the sky intensity distribution, we need to understand and model the primary beam of the instrument (i.e. $A(\hat{\mathbf{s}},\nu))$. This term determines the sensitivity and frequency dependence of the antenna and is dependent on frequency. The exact spatial and frequency structure, as well as polarization response, is complicated \citep{Ewall_Wice_2016, Patra2018, nunhokee, fagnoni}. The situation is more complicated for instruments like MWA and SKA with an array of dipoles, whose combined beam will produce the station primary beam. The complexities of some of the beam effects can be minimized through design specifications. However, they cannot be removed completely. The beam may also show variations amongst different feeds (or dipoles for tiled arrays) due to inherent structural/positional errors or malfunctioning electronics, which adversely affect EoR observations \citep{Joseph2020, Nasiruddin2022}. 

\subsection{Other Contaminants}
Radio interferometric data often suffer from the presence of RFI in frequency channels. RFI is sudden unwanted bright radio signals from different terrestrial sources. They are usually localized in frequency, affecting a few frequency channels. Radio astronomical observations are usually taken with high time and frequency resolution \citep{rfi1}, which helps identify RFI contamination and subsequently flag the affected timestamps/frequency channels. This introduces irregular visibility samples, which introduces spurious structures in the data. While performing Fourier transforms for calculating PS, these irregularities introduce oscillatory structures in the PS domain. Since the target signal is weak, this can adversely affect the data recovery\citep{rfi2}.

Low-frequency data are also affected by the earth's ionosphere. The incoming extra-terrestrial wavefronts suffer refraction from the non-uniformity of the ionospheric plasma layers. The presence of plasma gradients across the pierce points to the extragalactic sources results in offsets in source positions, while curvature can change the source flux density \citep{vedantham_iono}. At the frequencies of interest for CD/EoR science ($\lesssim$ 150 MHz), the ionospheric effects can become quite significant, affecting data calibration and, subsequently, signal recovery. 

However, despite the numerous factors affecting the recovery of the redshifted 21-cm signal, it is expected that the next generation radio telescopes can detect it statistically and even perform tomography. There is meticulous research ongoing to investigate the manifestation of these corruptions and their mitigation strategies. While detection is yet to be achieved, the operational instruments have all provided extremely sensitive upper limits on the signal PS. The following subsection describes the current results obtained from present observations.

\section{Current Status of 21-cm Observations: Power Spectrum}
\label{upper_interferometer}
The majority of the low-frequency instruments operating at $\lesssim$150\,MHz are actively targeting the detection of the power spectrum of the 21-cm signal. Thanks to the enhanced sensitivity and advanced instrumentation, coupled with the development of advanced data analysis and error mitigation techniques (already discussed above), there have been extremely sensitive upper limits on the \hi power spectrum from EoR and CD. The GMRT, LOFAR, MWA, and HERA have placed sensitive limits in the last decade. With the SKA already in the construction phase, and SKA-1 Low expected to push these limits even lower to achieve actual detection of the 21-cm PS, it is worthwhile to explore the current limits on the spherically averaged PS  ($\Delta^2(k)$) set by the most sensitive low-frequency instruments till date. We provide a brief description of the limits till date below, and the corresponding values are tabulated in Table \ref{upper}.

\subsection{GMRT}
The Giant Metrewave Radio Telescope (GMRT) \citep{Swarup_1991} is one of the largest and most sensitive fully operational low-frequency radio telescopes in the world, located at Khodad (close to Pune) in India. The array consists of 30 fully steerable parabolic antennae, of 45m  diameter each,  spanning over  25 km providing a  total collecting area of about  30,000 m$^{2}$  at meter wavelengths,  with a  fairly good angular resolution ($\sim$arcsec). The legacy system has been recently upgraded to the uGMRT with enhanced frequency coverage (120-1500\,MHz), wider bandwidth ($\sim$400\,MHz), and improved receiver systems - all contributing to better sensitivity and dynamic range \citep{Yashwant_2017}. 

GMRT was one of the pioneering instruments for CD/EoR science, placing the first upper limit on the 21-cm power spectrum at redshift $z\sim 8.6$ \citep{paciga2011}, using $\sim 40$ hours of observation with the legacy GMRT. The data was calibrated using a pulsar B0823+26 as a calibrator. Following calibration, RFI removal, and foreground subtraction, data from multiple observing nights were cross-correlated to set the upper limit. A revised upper limit, after accounting for signal loss due to foreground subtraction, was placed by \citep{paciga2013}. This was still the most stringent upper limit on EoR 21-cm signal at that time. In addition to these observations and the corresponding upper limits on the \hi\ 21-cm power spectrum at EoR redshift,  quite a significant amount of work have been done using GMRT at frequencies corresponding to the post-EoR epoch. The challenge of extremely bright foregrounds is same for both EoR and post-EoR \hi\ signal detection. Hence, the knowledge gathered from the post-EoR observation can be used to EoR observation and vice-versa. \citet{Ghosh_2011} used GMRT observations at 618 MHz to put the first upper limit on [$ \bar{x}_{\mathrm{HI}} b_{\mathrm{H{\sc I}}}$] $\leq$ 2.9 ([$\Omega_{\mathrm{H{\sc I}}} b_{\mathrm{H{\sc I}}}$] $\leq$ 0.11) at $z \sim 1.32$, where $ \bar{x}_{\mathrm{HI}}$ is the mean neutral fraction. The uGMRT data centred at 400\,MHz was used by \citet{arnab2021} to place the first upper limits on the post-EoR 21-cm PS in the range $2\lesssim z \lesssim4$. They also constrained [$\Omega_{\mathrm{H{\sc I}}} b_{\mathrm{H{\sc I}}}$] with upper limits 0.09,0.11,0.12,0.24 at $z=1.96,2.19,2.62,3.58$, respectively.

\subsection{MWA}
The Murchison Widefield Array (MWA)\footnote{\url{https://www.mwatelescope.org/}} \citep{mwa_phase1, mwa_phase2}, is a precursor to the SKA-Low, located in the Murchison Radio Observatory in Western Australia. It is an aperture array, operational between 80-300 MHz. At the initial phase, it consisted of 128 square ``tiles" of 4\,m$\times$4\,m, distributed over $\sim$3 km, which later got upgraded to 256 tiles over $\sim$5 km \citep{mwa_phase2}. The post-upgrade configuration has 72 tiles arranged into a highly redundant ``compact" hexagonal configuration for EoR science. This configuration improves the PS sensitivity and has helped it to place upper limits on the EoR signal.   

The initial upper limit using a 32-tile prototype for MWA was set by \citet{dillon_2015}. Their method employs a robust estimator that tracks error covariances and provides PS estimates using the k-modes in the EoR window. Using 128 tile MWA configuration,  \citep{mwa_l1} provided the upper limit between $z$ 17.9 and 11.6 or $\nu$ 75 \& 113 MHz \citep{mwa_l1} for 2 nights of observations. The systematic errors reported in \citep{mwa_l1} were mitigated to a large extent in \citet{mwa_l2}, and the techniques were employed to deep 32 hour integrated data, providing the upper limits at $z=$7.1. \citet{mwa_l3} further improved systematic conditions and observed effects, improving the PS limit using just 21 hours of data by almost an order of magnitude. Using the improved techniques developed in \citep{mwa_l3} along with some additional quality checks, 40 hours of MWA Phase-II data was used to provide upper limits at $z$ 6.5, 6.8, and 7.1 \citep{mwa_l4}. A multi-redshift limit on the 21-cm PS over the range $6.5\leq z \leq 8.7$ was set by \citet{mwa_l5}, using four seasons of observation from the MWA EoR project. Using $\sim$300 hours of cleanest data from the combined set, they provided PS estimates over different fields, pointings, and redshift ranges. Recently, \citet{mwa_l6} has also provided an upper limit on the 21-cm PS at $z\sim$13-17, using 15 hours of MWA data. The different values of the lowest value of $\Delta^2(k)$ provided by MWA so far have been tabulated in Table \ref{upper}.

\subsection{LOFAR}
The Low Frequency Array (LOFAR) \footnote{\url{https://www.astron.nl/telescopes/lofar/}} \citep{lofar_const} is a low frequency telescope operational in the range 30-240 MHz \citep{lofar_freq}. LOFAR, an SKA pathfinder, is a phased aperture array with tiles or stations spread over nine countries in Europe, centered in the Netherlands. It is an excellent low-frequency telescope that has so far set very sensitive upper limits on the \hi 21-cm PS out to $z \sim$25. 

The first upper limit from the LOFAR was obtained with $\sim$ hours of data from its High-Band Antenna, between $z$ 7.9-10.6 \citep{lofar_l1}. Limits on the Cosmic Dawn were obtained from 14-hour data of the Low Band Antenna between $z$19.8-25.2 \citep{lofar_l2}. The most sensitive limit obtained by LOFAR to date is obtained at a redshift of 9.7 using 141 hours of LOFAR data \citep{lofar_l3}. This provided almost an order of magnitude improvement over the previous limits. Through all the EoR PS analyses, the authors have reported the observation of ``excess variance"  in the observed PS, despite correcting for systematic effects and foreground removal. Recent investigations into the cause of this variance attributed it to sky-related effects, particularly related to distant bright sources \citep{excess_variance}.

\subsection{HERA}
The Hydrogen Epoch of Reionization Array (HERA) \footnote{\url{https://reionization.org/}} is a staged experiment to detect \hi 21-cm signal from CD/EoR in the redshift range 6-30 \citep{hera_const}. Located in the Karoo Radio Astronomy Reserve in South Africa, HERA will operate in the range 50-250\,MHz. It is currently under active construction and will consist of 350 dishes of 14m diameter each, spread out to $\sim$850 m,  once fully operational. 

The most sensitive limit on the \hi 21-cm PS till date is given by the HERA telescope \citep{hera_l1}. Using 18 nights ($\sim$36 hours) of observing time and using 39 of the 52 operational antennas, they achieved an order of magnitude improvement over the previous observations, placing limits at $z=$ 7.9 and 10.4.

\begin{table*} 
\begin{center}
\caption{Upper Limits on the \hi 21-cm Power Spectrum from CD/EoR set by different telescopes}
\label{upper}
\begin{tabular}[\columnwidth]{lcccl}
\hline
\hline
Telescope &  z & k & $\Delta^2(k)$  & Reference \\
& &(Mpc$\rm ^{-1}$ h) &(mK$^{2}$)& \\
\hline
&&&& \\
GMRT & 8.6 & 0.65 $h$ & $<4.9\times10^{3}$ & \citealt{paciga2011}\\
&&&& \\
& 8.6 & 0.5$h$ & $<6.1\times10^{4}$ & \citealt{paciga2013}\\
&&&& \\
& 1.96 & 1.0 & $<3.5\times10^{3}$ & \\  
& 2.19 & 1.0 & $<3.8\times10^{3}$ & \citealt{arnab2021}\\
& 2.62 & 1.0 & $<3.7\times10^{3}$ & \\
& 3.58 & 1.0 & $<1.1\times10^{4}$ & \\
\hline
&&&& \\
MWA & 12.2 & 0.18 & $<2.5\times10^{7}$ & \\
&  15.35 & 0.21 & $<8.3\times10^{7}$ & \citealt{mwa_l1}\\
& 17.0 & 0.22 & $<2.7\times10^{8}$ & \\
&&&& \\

& 7.1 & 0.27 & $<2.7\times10^{4}$ & \citealt{mwa_l2}\\
&&&& \\

 & 7.0 & 0.20 & $<3.9\times10^{3}$ & \citealt{mwa_l3} \\
&&&& \\

 & 6.5 & 0.59 & $<3.39\times10^{3}$ & \citealt{mwa_l4} \\
&&&& \\

& 6.5 & 0.142 & $<1.8 \times10^{3}$ & \\
 & 6.8 & 0.142 & $<3.6\times10^{3}$ & \\
 & 7.1 & 0.142 & $<6.0\times10^{3}$ & \citealt{mwa_l5}\\
 & 7.8 & 0.142 & $<2.4\times10^{4}$ & \\
 & 8.2 & 0.142 & $<2.8\times10^{4}$ & \\
 & 8.7 & 0.142 & $<6.2\times10^{4}$ & \\

 & 15.2 & 0.14 & $<6.3\times10^{6}$ & \citealt{mwa_l6} \\
&&&& \\
\hline
&&&& \\
LOFAR & 7.9-8.7 & 0.1 & $<1.1\times10^{5}$  & \\
 & 8.7-9.6 & 0.1 & $<8.7\times10^{4}$ & \citealt{lofar_l1}\\
 & 9.6-10.6 & 0.1 & $<5.0\times10^{4}$ & \\
&&&& \\
 & 19.8-25.2 & 0.038 & $<2.1\times10^{8}$ &\citealt{lofar_l2} \\
&&&& \\
 & 9.1 & 0.075 & $<5.3\times10^{3}$ &\citealt{lofar_l3} \\
\hline
&&&& \\
HERA & 7.9 & 0.192& $<9.5\times10^{2}$ & \citealt{hera_l1} \\
 & 10.4 & 0.256 & $<9.1\times10^{3}$ & \\
&&&& \\
\hline
\hline
\end{tabular}
\end{center}

\end{table*}

\section{Simulating Observations}
\label{sims}

As discussed in the previous subsection, numerous sources of corruption affect observations of the cosmological signal. Different studies have described how each of these factors affects the observations. However, systematics and foreground mitigation generally receive more attention since they are the primary contaminants. It was shown in \citet{Datta2009, Datta2010} that the presence of residual calibration errors, as low as $\sim$0.1\%, can make the residual foreground power leak from the wedge, contaminating the EoR window. Thus, the residual foregrounds dominate the PS, thereby suppressing the signal. They also find a similar power leakage and subsequent signal suppression when source position errors are $\sim$0.1". It was later shown in \citet{trott2012} that residual from source peeling for flu densities above 1 Jy would not affect the PS estimation significantly. \citet{thayagrajan2013} showed that a finite bandpass window makes unsubtracted point sources and their side lobes spill into the EoR window, affecting the PS. They also show that this leakage can be reduced to a large extent by switching to a Blackman-Nuttal window instead of a rectangular one. The foreground handling strategy -  avoidance or suppression also affects the data recovery. \citet{chapman2016} had performed a comparative study of the methods. They concluded that both avoidance and removal perform well for foreground models with line of sight and spatial variations of $\sim$0.1\%, but removal shows better recovery at even higher variations. 

It is thus seen that 21-cm observations are systematics limited. With the SKA promising detection of the coveted signal, it is imperative to handle how these systematics will affect SKA-Low observations. Hence, a consolidated pipeline that can simulate different realistic observational systematics is required. This will help understand the problems that are to be expected from the SKA-Low data and strategize their mitigation. With this goal, an end-to-end pipeline has been developed that can simulate realistic mock observations and incorporate different errors in the simulation to enable investigations into the effects. The following subsections describe the pipeline and demonstrate a few results obtained.

\subsection{Description of Pipeline}
\begin{figure*}

  \includegraphics[width=2\columnwidth, height=10cm]{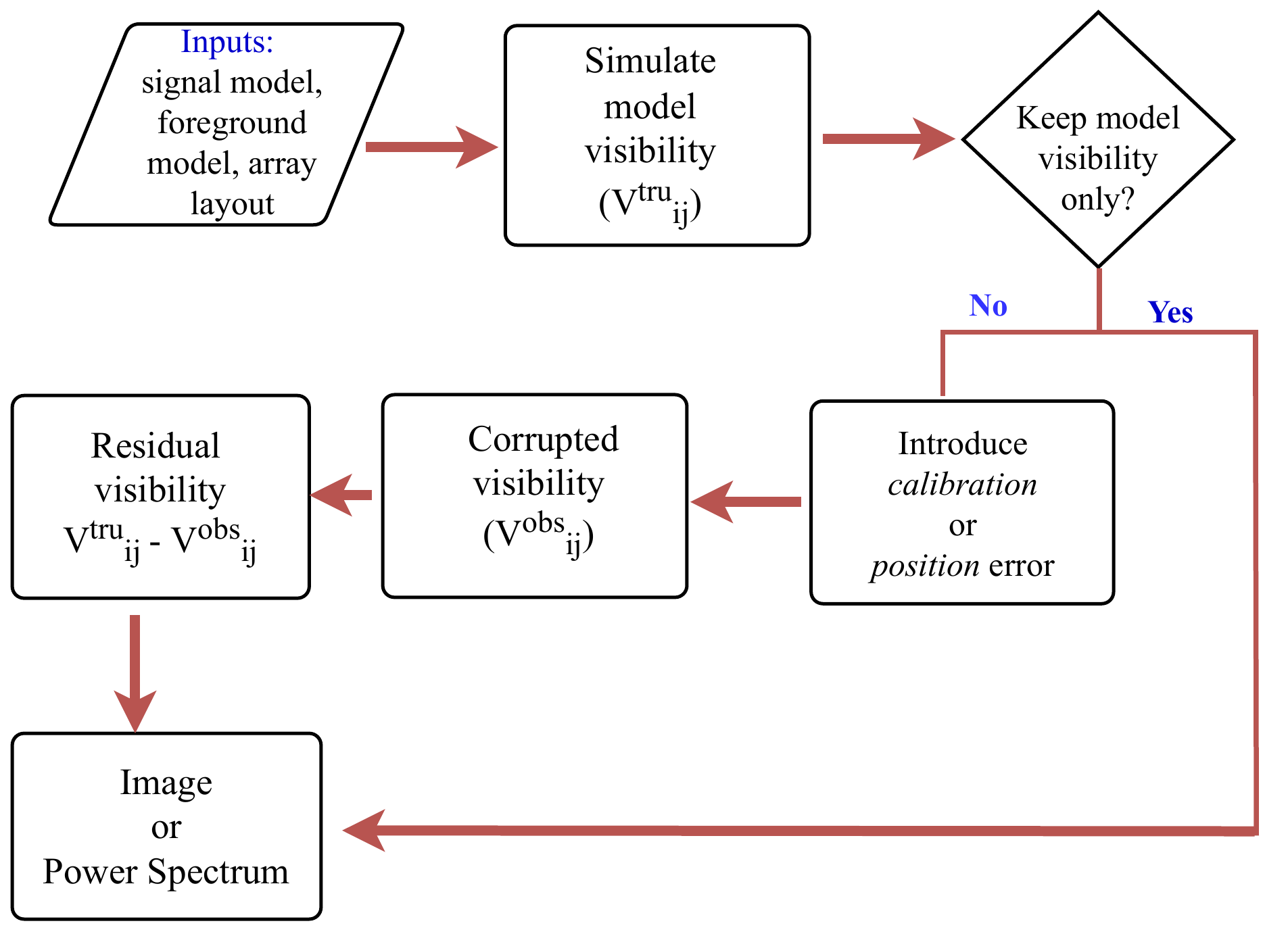}  
   \centering
    \caption{Flowchart describing the basic operation of the end-to-end simulation pipeline}
    \label{flow}
\end{figure*}

With the starting of the construction phase of the SKA-1, there is currently a requirement for performing mock observations under realistic conditions to forecast SKA performance. The SKA Indian consortium is actively contributing to the SKA through different aspects. We have designed an end-to-end simulation pipeline for mock interferometric observations as part of these ongoing efforts. The basic workflow is depicted in Figure \ref{flow}. 

In the current version, it is a composite mixture of the Common Astronomy Software Applications simulation tool (\casa \footnote{\url{https://casa.nrao.edu/}} \citet{casa}) and \osc \footnote{\url{https://github.com/OxfordSKA/OSKAR}} software package \citep{oskar}. \casa is one of the most widely used calibration and imaging software in radio astronomy. It is built principally using the \texttt{C++} language along with some \texttt{FORTRAN} with an \texttt{I{\sc Python}} interface that provides the capability of interferometer data viewing, calibration, editing, and imaging. There are in-built tasks for each step of data reduction. It also has inbuilt tasks for the simulation of data. Additionally, several toolkits are available for different applications, including data simulation. The \casa \emph{simulator} tool provides a large number of functionalities for simulations using different array layouts, frequency and time information, beam patterns, noise, etc. \osc is another simulation package that produces interferometric visibilities for aperture arrays. \osc is written principally in the C programming language and has GPU acceleration capability, which can be run from both bash and using {\sc Python} interface. It provides options to simulate the radio sky using source catalogues or fits files, array and array element layout files for aperture arrays like the SKA-Low, along with different beam patterns, noise properties, etc. A combined pipeline is thus capable of generating mock observations with both dish telescopes and aperture arrays.

As can be seen from Figure \ref{flow}, the first step in the simulation is providing the input files - namely array layout file, signal model, and/or foreground model. The layout file should have the coordinates of each antenna/station. For \casa, the local tangent coordinates are used, while \osc accepts both local tangent and WGS84 coordinates. Additionally,  aperture arrays like SKA have a set of dipoles in each station/tile (whose combined response is the ``station beam", equivalent to the primary beam for each telescope in case of a traditional dish array). Thus, \osc also requires separate layout files for each dipole in each station of the aperture array. The next input parameter is a model of the 21-cm signal. There are different approaches like radiative transfer \citep{finalator, artist}, semi-numerical methods \citep{21cmf, suman2014}, ray tracing \citep{ray_eor}, radiation magnetohydrodynamics \citep{mhd_eor} etc to generate the 21-cm ionization fields, brightness temperature distributions, perturbed fields etc. However, semi-numerical techniques remain the most popular method due to their computational efficiency and reasonable accuracy. Currently, we have tested the pipeline with two popular publicly available semi-numerical software for generating 21-cm signal - 21cmFAST\footnote{\url{https://github.com/21cmfast/21cmFAST}} \citep{21cmf, Murray2020} and ReionYuga\footnote{\url{https://github.com/rajeshmondal18/ReionYuga}} \citep{tirth2009, suman2014, reionyuga}. The brightness temperature of the 21-cm signal is generated with a specified box length, grid size, and redshifted range. To ``observe" this signal, it has to be associated with a world coordinate system. We calculate the angular size corresponding to the box size at the redshift of interest and the frequency range corresponding to the redshift range. Then using the RA, Dec of the phase center, a dummy image using the frequency and angular size information is generated, where the brightness temperature is projected (after conversion to Jy beam$^{-1}$). This is used as the input signal. For the point source foregrounds, a source catalogue containing RA, Dec, and flux density of the sources is required. Optionally, reference frequency, source size, polarized intensity, and spectral index can also be included. In case of \osc, this is a simple ASCII catalogue. But for \casa, one needs to generate a \emph{componentlist}, which is a source list in \casa specified format \footnote{See \url{https://casa.nrao.edu/docs/casaref/componentlist-Tool.html\#x237-2380001.2.1}}. Foreground model can include any source catalogue generated from observations as well as semi-analytical simulations like T-RECS \citep{trecs}. 

The signal and foreground models can also be used separately to simulate respective observations. In addition to the telescope layout and signal/foreground models, we specify the observing time, frequency, integration time, and optionally the primary beam and noise. 
The inputs mentioned above generate visibilities in the form of a measurement set \footnote{Measurement sets are the observed visibilities written in a particular format that is understood by \casa. For more information, see \url{https://casa.nrao.edu/Memos/229.html}}. This is step constitutes a complete synthetic observation. The pipeline has further capabilities for studying different systematic errors. Currently, we have implemented two sources of errors- calibration and position error. For calibration error, we corrupt the model visibilities by imperfect instrumental gains (Equation \ref{gains_eq}), generated per timestamp per antenna. These corrupted visibilities are subtracted from the model ones to produce residual visibilities. Position errors involve source displacement from its original position. Thus, it requires the generation of catalogues containing some systematic displacements of the sources from the real position. A new synthetic observation using the corrupted source catalogues produces corrupted visibilities. They are again subtracted from the real ones, producing residual visibilities. 

For studying the effects of these corruptions on the 21-cm signal recovery, the visibilities can either be imaged and post-processed or used directly for PS estimation. Imaging is performed using either \casa or WSCLEAN \citep{wsclean}. These are standard imaging software used for imaging radio astronomical data. For power spectrum determination, we use a PS pipeline developed based on the delay spectrum technique \cite{Parsons2012} and implements Equation \ref{2dps_eq}, which is further averaged according to Equation \ref{3d_ps}, producing the 2D and 1D PS. The spherical averaging can be done using either all the k-modes or only those outside the wedge. In the presence of RFI flagging, it implements either CLEAN \citep{Hogbom1974} or LSSA \citep{Patil2016} algorithms to reduce oscillatory features introduced due to non-uniform data samples. This has been used in \citet{arnab2021} to provide upper limits on the post-EoR PS and \citet{2022_arnab} to provide a comparison between the efficacy of different RFI mitigation methods. A systematic analysis of the performance of the whole pipeline along with the PS estimations has been presented in \citet{am_simulation}. They have compared the impact of systematics for signal recovery using different arrays under identical conditions of sky and foregrounds. The PS pipeline is currently being integrated with the main pipeline and will be publicly released after successful integration and testing. 

\section{End-to-end Pipeline at Work}
\label{perf_pipe}
\subsection{Demonstration of Pipeline Performance}

\begin{figure*}
    \includegraphics[width=\columnwidth]{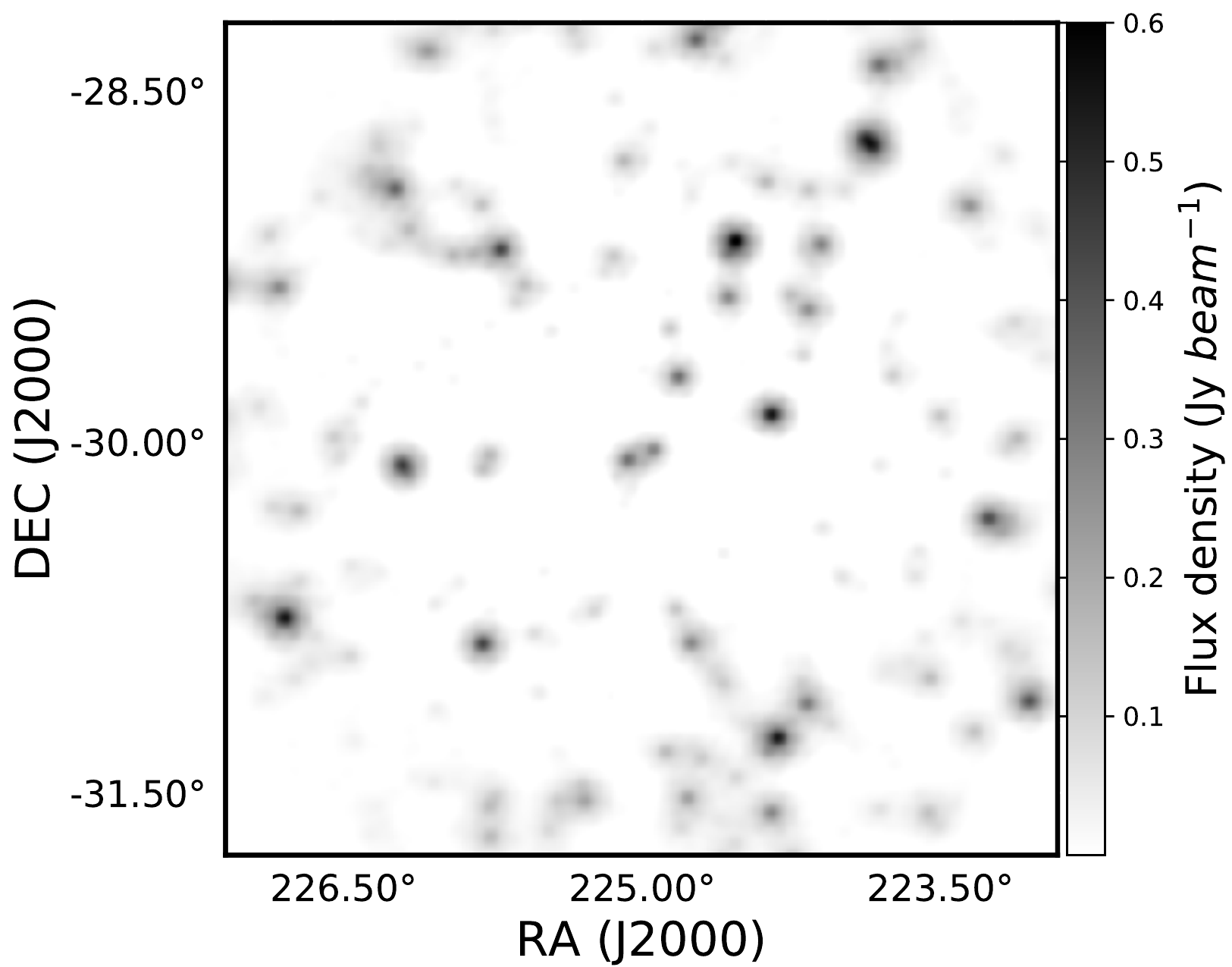}
    \includegraphics[width=\columnwidth, height=7cm]{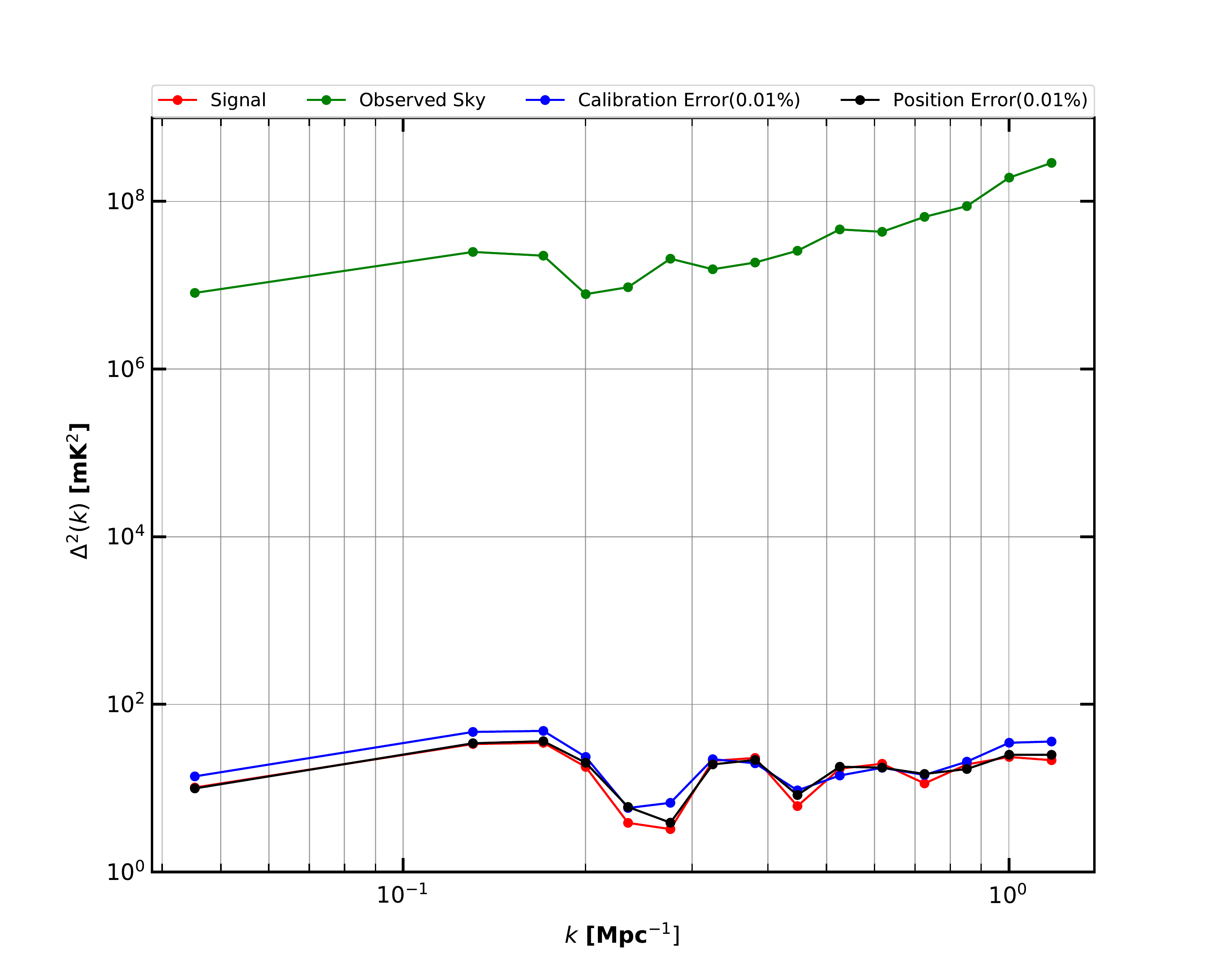}
    \caption{\emph{(left)} The simulated sky with SKA. \emph{(right)} The 1D power spectrum for the sky (signal+foreground, green), pure observed signal power(red), residual power with 0.01\% gain error (blue) and 0.01\% position error(black).}
    \label{demo_pipe}
\end{figure*}

Here, we demonstrate the performance of the end-to-end pipeline briefly. We have run a 10-minute snapshot observation with the SKA-1 Low station layout located within $\sim$2 km of the central station\footnote{The latest SKA-1 layouts are available at \url{https://astronomers.skatelescope.org/DOCUMENTS/}}. The sky-model used is a 21-cm map generated from ReionYuga and point source obtained from the 400 MHz uGMRT observation of the ELAIS-N1 field \citep{arnab2}. The simulation was done for a 32 MHz bandwidth, spread over 64 frequency channels centered at 142\,MHz\,($z\sim$9). These are the model visibilities, which we corrupt with 0.01\% calibration error and position error of $\sim$0.02" (0.01\% of the PSF size). For most science cases of interest, a calibration error of$\sim$10\% is tolerable. However, it was shown in \citet{Datta2009, Datta2010} that for sensitive observations targeting the EoR signal errors $\gtrsim$0.05\% cause the residual foregrounds override the target. Thus, following that work, the test case taken here has 0.01\% calibration error introduced. For position errors, while \citet{Datta2009, Datta2010} provide a tolerance threshold of $\sim$0.1". But more detailed simulation of \citet{am_simulation} showed a tolerance of about $\sim$ a few arcseconds. However, the test case shown here considers an even lower position inaccuracy (which may be possible with the large baselines and sensitive design for the upcoming SKA). We obtain the residual visibility by subtracting the model from the corrupted ones (see Figure \ref{flow}) and calculate the spherically averaged PS from it. The results are shown in Figure \ref{demo_pipe}. It is seen that for an error of 0.01\% in gain calibration, or 0.01\% in source position, the \hi power spectrum is recoverable. 

The simulations have been performed in a GPU enabled machine with a mere 64 GB RAM capacity. \osc is able to run on both GPU and CPU. For a typical 4 hour run with the frequency the specifications described above, the simulated data volume is about $\sim$3\,GB. The simulation time for the test cases with aforementioned specifications in \osc is $\sim$2.5 hours while in CPU it takes around 6 hours. Thus, GPU improves the run times by a factor of 2.4. On the other hand, \casa is not GPU enabled, and a typical run takes between 8-12 hours depending on the array, i.e. the number of interferometer elements used. The major drawback for \osc is that is works only for aperture arrays like the SKA and not for dish arrays like HERA.

This is a very simplistic demonstration of the working of the pipeline. A detailed analysis has been presented in \citet{am_simulation}. Later papers will also demonstrate the improvements as they happen.

\subsection{Parameter Estimation}
\label{pe}
The ultimate goal of any observation of Global signal/PS measurements is parameter estimation, i.e., constraining the physical parameters controlling the astrophysics and cosmology influencing the IGM and its ionizing sources. The past decades have seen meticulous research in this regime in conjunction with foregrounds and systematics. The most popular method remains Bayesian statistics, but with the advent of machine learning (ML), there has been a significant focus on using ML as well. This section briefly discusses the current state of the art in parameter estimation.

The Markov Chain Monte Carlo (MCMC) is the most commonly used method to estimate the best-fit parameters from the multidimensional parameter space of cosmological data. The posterior probability distribution is created using the Bayesian approach. For sampling the posterior, MCMC algorithms used one of the most widely used samplers called the Metropolis-Hastings (MH) \cite{Metropolis, Montecarlo} sampler. The MH algorithm draws the sample based on a random walk over the likelihood of the parameter space by serially proposing new accepted or rejected locations based on the likelihood weights. The MH algorithm generally considers the target distribution and the proposal conditional distribution from which a candidate sample for the new Markov chain state is drawn. The disadvantage of using MH samplers is that they typically require several thousands of model evolutions, and only a small portion gets accepted. Hence, applying the algorithm to problems where the model evaluation is computationally time-consuming is challenging. The intrinsic serial nature of the MH chains often takes a long time to map the posterior. Although MH delivers the intended result, it may struggle to do so efficiently if a suitable proposal function is challenging to discover or if the assumption of a fixed proposal function proves to be inefficient. Several other sampling algorithms are available, which may be more robust than the MH algorithms, for example, Slice sampling \citep{neal2003slice}, Gibbs sampling \citep{gelman1995bayesian, george1993variable, gelfand2000gibbs}, Hamiltonian sampling \citep{leimkuhler2004simulating,neal2011mcmc} .

Bayesian inference-based methods are intensive in computation and execution time. Therefore, new techniques are required to analyze the enormous amount of data produced by the operational and upcoming telescopes. ML algorithms are thus designed to extract correct theoretical predictions from training sets (instead of thorough computations of the physical equations at each parameter point). ML methods can efficiently learn from very complex, non-linear, high-order, non-Gaussian priors. The main advantage of using ML is that it trains the program to learn from their experiences and improve themselves. In supervised learning methods, the learning process usually involves optimization of a `cost function' or an `error function' by a technique called back-propagation. Unsupervised learning methods can be used on complex data sets to find different correlations present in the data. There are many other advantages of using unsupervised learning methods as they are capable of self-learning, self-adaption, robustness, and have a dynamically rapid response. Several different machine learning algorithms have been used in various elements of astronomy, astrophysics, statistics and inference, and imaging throughout the last several years. For example, \citet{schmit2018emulation} created an emulator for the 21-cm power spectrum that used artificial neural networks (ANN). This emulator was also utilized within a Bayesian inference framework to restrict the EoR parameters using the 21-cm power spectrum. \citet{cohen2020emulating} used ANN to create an emulator for the 21-cm Global signal, linking astrophysical parameters to the projected global signal. \citet{hassan2019identifying} used Convolutional Neural Networks (CNN) to identify different reionization sources from 21-cm maps. In \cite{chardin2019deep}, deep learning models were used to simulate the full time developing 21-cm brightness temperature maps from the period of reionization, and the authors compared their anticipated 21-cm maps to the brightness temperature maps obtained by radiative transfer4964.18 simulations. \citet{Paul_2019} used CNN to recover the duration of reionization from reionization images, assuming that the midpoint of reionization is already well constrained. \citet{gillet2019deep} used deep learning and CNN to retrieve astrophysical parameters directly from 21-cm pictures. Machine learning strategies for estimating the 21-cm power spectrum from reionization simulations were compared by \citet{jennings2019evaluating}. By training on SKA pictures produced with realistic beam effects, \citet{Li(2019)} constructed a convolutional de-noising autoencoder (CDAE) to recover the EoR signal. \citet{Zhao_2022} applied CNN to 3D-tomographic 21-cm images for parameter estimation and posterior inference.

The preceding examples focus on parameter estimation in the image domain. In contrast, \citet{Shimabukuro_2017} have used ML algorithms for extracting the parameters of the 21-cm power spectrum; however, they have not considered foregrounds in their analysis. This has gap has been addressed in \citet{choudhury2020extracting}, who employed ANN to extract parameters from a 21-cm Global signal under more realistic conditions. They used bright galactic foregrounds and instrumental corruption to estimate astrophysical parameters at modest computational costs. Their parameter estimation method gave $\gtrsim$92\% accuracy, even for corrupted data sets. In \citet{choudhury2021using}, they have developed ANN models to directly extract astrophysical characteristics from 21-cm Global signal observations, using physically justified 21-cm signal and foreground models. They have further developed their network in \citet{mc3} to extract the 21-cm PS and the corresponding EoR parameters from synthetic observations. There are currently efforts to include more realistic effects like ionosphere and primary beam in the training process to produce networks that can predict parameters from corrupted data sets (\textit{Tripathi et al. in prep}). Once perfected, these would be incorporated into the pipeline described above for recovering parameters from the synthetic observations using realistic conditions.

\subsection{Limitations}
\label{limit}
The end-to-end pipeline demonstrated above synthesizes interferometric observations using realistic foreground and systematic models. While it can be used for 21-cm interferometric observations and other science cases if required, there are certain limitations to it. These are as follows:
\begin{itemize}
    \item \emph{Diffuse Foregrounds}: In its current form, the pipeline does not incorporate any method to include the DGSE. There is a huge database of diffuse emissions available at the Legacy Archive for Microwave Background Data Analysis (LAMBDA\footnote{\url{https://lambda.gsfc.nasa.gov/}}). These consist of observations at different frequencies by different telescopes and all-sky simulations. Our simulation pipeline is currently missing the capability to incorporate these maps into mock observations and methods for their mitigation. This will be implemented, and later publications will demonstrate the results. 
    
    \item \emph{Noise}: Thermal noise is also a systematic error that needs to be incorporated for making simulations more realistic. There are options to implement them in both \casa and \osc. However, that has not been done for the initial work. We are currently incorporating noise in our simulation to determine its effect in the presence of other error sources. 

    \item \emph{Computational Requirements}: The \casa implementation requires about 3$\times$ the \osc implementation since it is purely CPU based (the latter has both CPU and GPU capabilities). Thus, there is a requirement for a faster implementation, which will be explored in the future. Additionally, the PS pipeline requires a sizeable memory $\sim$128\,GB for a mere 4-hour mock observation with the SKA core in total intensity (Stoke's I) only. Thus, it requires HPCs for analysis. 
    
    \item \emph{Parameter Estimation}: The parameter estimation methods under development, as described in Section \ref{pe} is a separate development and is yet to be tested on synthetic data obtained from our end-to-end simulation. 
\end{itemize}

It should be mentioned here that the pipeline is an effort to provide the SKA community with a simulation tool for checking the implications of realistic observation conditions. While this is developed keeping EoR science in mind, it can easily be extended for other science cases. It is a work in progress, and we plan to release it publicly soon. 

\section{Summary}
\label{summary}
This review explores the practical aspects of detecting the redshifted \hi 21-cm signal using radio telescopes. With the SKA having begun its construction phase, the time is suitable for perfecting observational strategies and data analysis methods to get the best results once the SKA starts taking data. This involves understanding the target and the factors that hinder its detection. The signal of interest has several factors controlling its nature, strength, and time evolution. This comes mixed with the bright foregrounds in the intervening path and different instrumental influences in the telescope. Thus, we need to understand each of these aspects and their interplay for successful signal recovery. Keeping these requirements in mind we have developed an end-to-end pipeline for synthesizing interferometric observations under realistic considerations of the sky and systematics. This review aims at highlighting the current status of this indigeneously developed simulation pipeline.

We have started with a brief overview of the observable parameter for the 21-cm signal, namely, its differential brightness temperature. Observation of the signal is done through brightness contrast against the CMB. This provides two ways to detect it - via all-sky averaged ``global" signal or statistically through fluctuations in the brightness temperature (PS). The basic principle behind each of these methods has been mentioned. The dominant signal suppressant - astrophysical foregrounds have also been explored, and their influence and possible mitigation strategies have been mentioned. Different instrumental systematics affecting global signal and PS observations have also been described. Despite the enormous number of obstacles, there has been a lot of observational headway in the past decade. The EDGES telescope has reported an apparent trough in the global brightness temperature spectrum at 78 MHz ($z\sim$17), where theoretical models predict the presence of a similar feature \citep{EDGES2018}. The depth and width of the detected profile, being deeper and broader than predicted, remains a topic of much research. Nevertheless, this was a significant result in the observational front. On the other hand, the most sensitive operational radio interferometers have placed meaningful upper limits on the signal PS from EoR \citep{paciga2011, paciga2013, mwa_l5, lofar_l3, hera_l1} and post-EoR. These limits have helped constrain astrophysical and cosmological parameters at play during the early epochs (for example, see \citet{2022ApJ...924...51A, arnab2021}). 

The current stage of the pipeline is discussed. In its present state, the pipeline can simulate the 21-cm signal and foregrounds, either individually or combined, from sky maps and/or catalogues. It can also incorporate realistic systematics and determine PS from the image or visibility domain. The pipeline is still under active development and is being tested for different systematics and diffuse foregrounds. Finally, we have discussed parameter estimation from observations. We describe existing methods like Bayesian methods and ML- the best bet for parameter estimation from large data sets. The focus of the simulation pipeline during the initial stages was for EoR science. However, this is a general interferometric simulation pipeline. Thus it will be helpful to the entire SKA user community, irrespective of the science goals.

\section*{Acknowledgment}
AM would like to thank Indian Institute of Technology Indore for supporting this research with Teaching Assistantship. AM further acknowledges Sarvesh Mangla, Swarna Chatterjee and Sumanjit Chakraborty for helpful discussions and moral support. AD would like to acknowledge the support from CSIR through EMR-II No. 03(1461)/19.

\bibliography{ref}

\end{document}